\documentclass[lettersize,journal]{IEEEtran}

\usepackage{cite}
\usepackage{amsmath,amssymb,amsfonts}
\usepackage{graphicx}
\usepackage{subcaption}
\usepackage{pgfplots}
\usepackage{booktabs}

\usepackage{xcolor}      
\usepackage{longtable}   

\usepackage[hidelinks]{hyperref} 
\usepackage[capitalise]{cleveref}

\title{Integrated Noise and Safety Management in UAM via A Unified Reinforcement Learning Framework}

\author{Surya Murthy\thanks{Graduate Research Assistant, Oden Institute for Computational Engineering and Sciences, The University of Texas at Austin}, Zhenyu Gao\thanks{Postdoctoral Fellow, Department of Aerospace Engineering and Engineering Mechanics, The University of Texas at Austin.}, John-Paul Clarke\thanks{Professor and Ernest Cockrell, Jr. Memorial Chair in Engineering, Department of Aerospace Engineering and Engineering Mechanics, The University of Texas at Austin}, Ufuk Topcu\thanks{Professor and Judson S. Swearingen Regents Chair in Engineering, Oden Institute for Computational Engineering and Sciences, The University of Texas at Austin,  IEEE Fellow}}

\begin{document}

\maketitle

\begin{abstract}

Urban Air Mobility (UAM) envisions the widespread use of small aerial vehicles to transform transportation in dense urban environments. However, UAM faces critical operational challenges, particularly the balance between minimizing noise exposure and maintaining safe separation in low-altitude urban airspace, two potentially conflicting objectives that are often addressed separately. We propose a reinforcement learning (RL)-based air traffic management system that integrates both noise and safety considerations within a unified, decentralized framework. Under this scalable air traffic coordination solution, agents operate in a structured, multi-layered airspace and learn altitude adjustment policies to jointly manage noise impact and separation constraints. The system demonstrates strong performance across both objectives and reveals tradeoffs among separation, noise exposure, and energy efficiency under high traffic density. Among the three objectives, safe separation is accorded the highest priority, whereas the relative significance of noise and energy varies by location and is contingent upon financial and public policy considerations. The findings highlight the potential of RL and multi-objective coordination strategies in enhancing the safety, quietness, and efficiency of UAM operations.

\end{abstract}

\section{Introduction}

\IEEEPARstart{T}{he} urban air mobility (UAM) system has the potential to revolutionize urban transportation and support various urban services, including passenger mobility, cargo delivery, infrastructure monitoring, public safety, and emergency response. By utilizing electric vertical takeoff and landing (eVTOL) aircraft, electric short takeoff and landing (eSTOL) aircraft, and unmanned aerial vehicles (UAVs) at lower altitudes, UAM is expected to enhance public welfare, particularly in underserved local and regional communities. While UAM offers substantial opportunities for enhancing urban services and promoting public welfare, challenges such as community noise acceptance, safety concerns, and infrastructure requirements hinder its successful integration into existing urban environments~\cite{barsotti2024eco}.

Since UAM operates at altitudes significantly closer to urban residents than traditional air transport systems, it is essential to address the noise generated by UAM aircraft, as it may pose health risks, including sleep disturbances and an increased incidence of cardiovascular issues for individuals residing near flight operations. To mitigate aerial vehicles' community noise impact, the implementation of noise-aware flight operations planning is imperative, irrespective of the available aircraft noise reduction technologies~\cite{gao2024noise}. On the other hand, an air mobility system, like all aviation systems, is safety-critical in nature. Therefore, it is essential to maintain safe separation between aircraft to prevent collisions and ensure system safety~\cite{chen2024integrated}. It is clear to practitioners and policymakers that a future UAM system must effectively address both noise and safety concerns to ensure the system's success and sustainability.

Because of the distinct nature of managing noise and safety concerns in UAM, existing studies have developed different frameworks to address these two objectives separately. UAM noise control typically relies on centralized planning approaches, such as deterministic optimization, to address problems like noise-aware flight trajectory optimization~\cite{bian2021assessment,gao2023noise} for managing single-event noise, and air traffic flow management~\cite{gao2024noise} for managing cumulative noise. UAM safety assurance methods include both centralized and decentralized approaches. Traditional air traffic management approaches often rely on deterministic algorithms for conflict detection and resolution across multiple time horizons and can achieve globally optimal solutions. However, they quickly become infeasible for large systems due to extended computation times and lack robustness against single-point failures and disturbances~\cite{waltz2024self}. In decentralized approaches, each aircraft independently selects actions based on its local information and makes online decisions, enhancing the system's computational efficiency and resilience. Therefore, decentralized safety assurance approaches, such as employing reinforcement learning (RL) to dynamically adjust aircraft speeds and understand complex interactions between vehicles, have emerged as a significant area of research~\cite{DENIZ2024100157, brittain2020deep, zhao2021physics}.

So far, due to the inherent differences and potential conflicts between UAM noise control and safety assurance, very few existing works have concurrently addressed both important aspects. Considering the importance of managing both issues and the advantages of decentralized methods in air traffic management (ATM), this work aims to develop a unified decentralized ATM framework based on RL for managing both noise and safety issues in UAM. We formulate the problem as a multi-agent reinforcement learning (MARL) problem in which multiple aircraft interact in a shared environment. For learning and control, each aircraft is modeled from its own perspective as an agent acting in a Markov decision process (MDP), where nearby aircraft are represented through local observations and the resulting interactions are captured through the environment dynamics and reward function. In this framework, we incorporate community noise exposure and loss of separation (LOS) events as objectives in the reward function. The RL agent learns to balance these objectives by adjusting altitude in structured airspace to mitigate ground noise and maintain safe vertical separation from other aircraft, thereby reducing the likelihood of horizontal LOS. We find that a trained RL agent can use altitude adjustments to reduce ground noise while maintaining safe separation between aircraft. Our results also highlight important tradeoffs among noise reduction, aircraft separation, and energy consumption. Overall, these findings demonstrate the potential of RL in managing multiple objectives in ATM and underscore the need for carefully balanced policies in UAM operations to achieve both environmental sustainability and operational safety and efficiency.

The remainder of this paper is organized as follows: In Section~\ref{sec:lit}, we review the literature across three relevant streams. Section~\ref{sec:method} details the proposed methodology, including aircraft performance models and the reinforcement learning model. In Section~\ref{sec:exp}, we describe the experimental setup using the South Austin UAM network as a case study. Section~\ref{sec:results} presents the main results of this study, followed by the conclusion in Section~\ref{sec:conclusion}.

\section{Background and Literature Review}\label{sec:lit}

\subsection{UAM Safety Management}

Safety is one of the highest priorities that UAM operators must uphold. Conflict resolution and separation assurance methods are essential for ensuring air traffic safety by identifying and resolving potential conflicts between aerial vehicles. In the aviation industry, systems such as ACAS X (Airborne Collision Avoidance System X) have employed advanced computational techniques such as probabilistic models for predicting aircraft trajectories and dynamic programming for generating avoidance maneuvers~\cite{manfredi2016introduction}. Within the UAM field, researchers have examined both strategic and tactical separation management approaches to facilitate safe operations.

Strategic separation management uses strategic decisions like ground delays made by air traffic managers to balance traffic demand with airspace capacity at bottlenecks. 
These bottlenecks include airport runways, merging points, and air route intersections.
Strategic separation management plays an important role in modern air traffic management, with programs such as the Ground Delay Program~\cite{odoni1987flow} and Airspace Flow Program~\cite{libby2005operational} serving as key examples.
Demand-capacity balancing (DCB) approaches use optimization and heuristic techniques to schedule flight takeoffs to balance demand with available capacity at bottlenecks~\cite{chen2024integrated}.
By integrating with real-time deconfliction, strategic separation approaches effectively enforce safety constraints while maintaining operational efficiency.

Tactical separation approaches address real-time conflict resolution by enforcing safe separation between aircraft, often utilizing techniques like Monte-Carlo Tree Search \cite{wu2022safety} and A* path planning \cite{zhao2021multiple}. Recently, researchers have made significant progress in applying deep reinforcement learning (DRL) methods for tactical separation assurance. DRL approaches model the UAM environment as a MDP and train aircraft to avoid conflicts using only local observations. Various DRL models incorporate advanced architectures, including long short-term memory (LSTM) networks \cite{DENIZ2024100157}, attention layers \cite{brittain2020deep}, and physics-based models \cite{zhao2021physics}. These advancements highlight the potential of DRL to enhance the safety and efficiency of air traffic management in UAM. 
Recent cooperative multi-agent reinforcement learning approaches such as MAPPO \cite{yu2022surprising} have also been applied to large-scale multi-agent coordination problems. Our work adopts a PPO-based learning framework similar in spirit to these methods, but differs in that both the actor and critic operate on local observations and neighboring aircraft information is aggregated using the attention-based D2MAV-A architecture \cite{brittain2020deep}.

\subsection{UAM Noise Control}

While UAM has significant potential for improving urban services and public welfare, its integration into existing urban environments presents several complex challenges. Among the operational constraints of UAM, aircraft noise has been identified as one of the most significant challenges during UAM development~\cite{vascik2017constraint,vascik2018analysis,pons2022understanding}. Since UAM operates near urban populations, the noise produced by UAM aircraft presents considerable health risks, including psychological distress, sleep disturbances, and a higher likelihood of cardiovascular diseases~\cite{gao2022multi}. Consequently, reducing UAM noise has become a key priority within the aviation and aerospace sectors.

Investigations into UAM noise mitigation strategies are grounded in two primary research pillars: aircraft noise modeling and noise-aware operational planning. Currently, UAM aircraft noise modeling mainly relies on acoustic simulations, primarily due to the lack of real-world measurement data for new eVTOL aircraft configurations~\cite{rizzi2022prediction,rizzi2023modeling,bian2021assessment}. Implementing noise-aware flight operations planning is essential for reducing the community noise impact of UAM, as it is applicable irrespective of the aircraft noise reduction technologies available. Optimization plays a key role in low-noise aircraft operations planning, including trajectory optimization and air traffic flow management. In commercial aviation, noise abatement trajectory planning strategies such as the continuous descent approach (CDA)~\cite{clarke2004continuous} and noise abatement departure procedures (NADP)~\cite{lim2020noise} have been the subject of extensive research. In the UAM sector, most research efforts have focused on generating noise-aware 2-D or 3-D flight trajectories for eVTOL aircraft and small drones~\cite{pang2022uav,tan2024enhancing}, as well as managing noise through airspace management strategies \cite{bauranov2021designing,gao2023noise,gao2025developing}. Air traffic flow management is effective in controlling the cumulative noise generated by UAM over specific time periods. By managing air traffic flow within the designated and networked flight corridors, it is possible to achieve limited cumulative noise exposure in urban areas \cite{gao2024noise}.

\subsection{Research Gap} 

Noise control and safety assurance are two critical objectives in UAM operations. However, very few existing approaches have sought to manage both aspects within the same framework. By examining current solutions, it is evident that centralized planning approaches, such as optimization, play a key role in UAM noise control, while decentralized control methods, such as DRL, are crucial for UAM safety management. Centralized approaches usually plan the entire trajectory of aircraft or air traffic flow assignments within a network, achieving globally optimal solutions. However, they can quickly become infeasible for large-scale systems due to extended computation times \cite{waltz2024self}. In contrast, decentralized methods allow each aircraft to make online decisions independently based on its local information. These approaches are computationally efficient and demonstrate greater resilience to single-point failures or disruptions.

Given (1) the need to concurrently manage UAM's noise and safety issues and (2) the advantages of decentralized methods in air traffic management, a natural question is ``Is it possible to manage both UAM's noise control and safety assurance within the same decentralized framework?'' To address this research question, we propose a DRL approach for the integrated management of UAM noise and safety. To the best of our knowledge, this represents the first pure RL solution that manages UAM noise impact while simultaneously considering separation assurance within the same framework. While most separation assurance approaches manage air traffic in a 2D plane and enforce planar separation between aircraft, our work allows aircraft to adjust altitude between multiple flight levels (cruising altitudes). This setup allows aircraft to decrease the likelihood of general LOS events by maintaining vertical distance separation and to mitigate ground noise impact by operating at higher altitudes. Each aircraft agent can dynamically adjust its altitude to minimize noise exposure while ensuring safe separation. 

\section{Methodology}\label{sec:method}

\begin{table}[htbp]
    \centering
    \caption{Key abbreviations used in Section~\ref{sec:method}}
    \begin{tabular}{ll}
    \hline
    \textbf{Abbreviation} & \textbf{Meaning} \\ \hline
    UAM   & Urban Air Mobility \\
    RL    & Reinforcement Learning \\
    DRL   & Deep Reinforcement Learning \\
    MARL  & Multi-Agent Reinforcement Learning \\
    MDP   & Markov Decision Process \\
    PPO   & Proximal Policy Optimization \\
    NPD   & Noise-Power-Distance \\
    SEL   & Sound Exposure Level \\
    LOS   & Loss of Separation \\
    D2MAV-A & Deep Distributed Multi-Agent Variable with Attention \\
    \hline
    \end{tabular}
    \label{tab:abbreviations_method}
\end{table}

\begin{table}[htbp]
    \centering
    \caption{Key variables used in Section~\ref{sec:method}}
    \begin{tabular}{ll}
    \hline
    \textbf{Variable} & \textbf{Description} \\ \hline
    $s_t$ & Ownship state at timestep $t$ \\
    $h_t^{(i)}$ & Observed state of neighboring aircraft $i$ \\
    $z$ & Current altitude of the ownship aircraft \\
    $z_{\text{target}}$ & Target altitude of the current maneuver \\
    $z_{\text{rel}}^{(i)}$ & Relative altitude of neighbor $i$ \\
    $b_{\text{changing}}$ & Indicator of whether an altitude change is in progress \\
    $a_t$ & Action taken at timestep $t$ \\
    $a_{t-1}$ & Previous action of the ownship aircraft \\
    $a_{t-1}^{(i)}$ & Previous action of neighboring aircraft $i$ \\
    $\mathcal{N}_{\text{adj}}$ & Number of altitude increases made so far \\
    $d_o^{(i)}$ & Horizontal distance between ownship and neighbor $i$ \\
    $d_{\text{comm}}$ & Communication radius for observing neighbors \\
    $d_{\text{LOS}}$ & Loss-of-separation threshold \\
    $R_{\text{noise}}$ & Noise reward component \\
    $R_{\text{separation}}$ & Separation reward component \\
    $R_{\text{energy}}$ & Energy reward component \\
    $\rho_{\text{noise}}$ & Weight on the noise reward \\
    $\rho_{\text{sep}}$ & Weight on the separation reward \\
    $\rho_{\text{energy}}$ & Weight on the energy reward \\
    $N_{\text{Single}}$ & Single-event noise level \\
    $N_{\text{increase}}$ & Cumulative noise increase above ambient level \\
    $N_{\text{ambient}}$ & Ambient community noise level \\
    $C_i$ & Weighted congestion around aircraft $i$ \\
    $C_{\text{max}}$ & Congestion normalization constant \\
    $\gamma$ & Discount factor in the MDP \\
    $\pi$ & Policy mapping states to actions \\
    \hline
    \end{tabular}
    \label{tab:variables_method}
\end{table}

\subsection{Aircraft Noise Model}

The aircraft noise model is a crucial component in noise-aware UAM operations planning. The aircraft noise modeling approach depends on factors such as range, fidelity, computational time, and urban terrain. The assessment of UAM aircraft noise impact on urban communities commonly relies on a norm called noise-power-distance (NPD) data~\cite{gao2024noise}. NPD data describes the relationship between an aircraft's noise impact and its slant distance to the receiver under different conditions. NPD data is currently available for approximately 300 fixed-wing aircraft and 26 helicopter types worldwide, derived from noise certification or controlled tests following strict international standards~\cite{volpe2022uam}. Due to the lack of NPD data for new aircraft configurations like eVTOL aircraft, practitioners rely on acoustic simulations to model noise patterns (e.g., 3D directivity) for these new configurations~\cite{rizzi2022prediction}. The simulation results are then converted into NPD data to effectively model UAM's community noise exposure.

\begin{figure}[h!]
     \centering
     \begin{subfigure}[b]{0.325\textwidth}
         \centering
         \includegraphics[width=\textwidth]{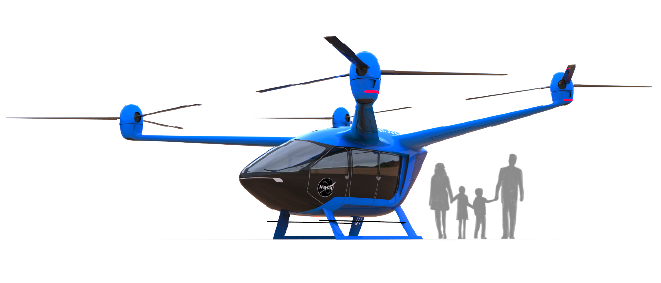}
         \caption{The NASA RVLT aircraft configuration}
         \label{fig:nasavehicle}
     \end{subfigure}
     \hspace{0.5cm}
     \begin{subfigure}[b]{0.375\textwidth}
         \centering
         \includegraphics[width=\textwidth]{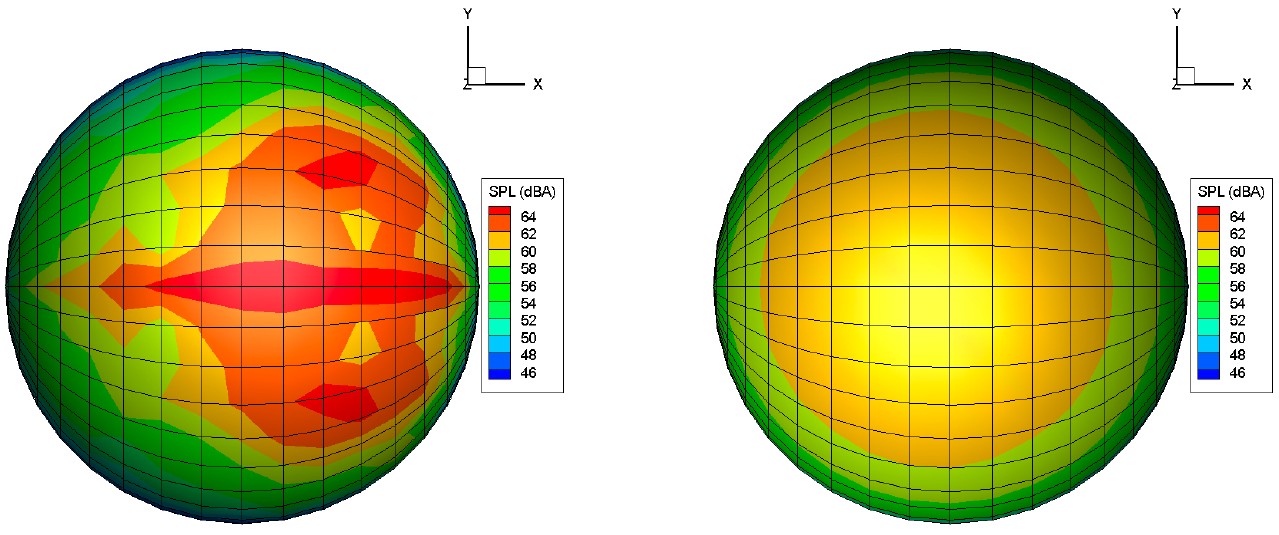}
         \caption{The noise hemispheres from acoustic simulation}
         \label{fig:noisesphere}
     \end{subfigure}
        \caption{The reference aircraft configuration and its noise hemispheres from simulation (sources: \cite{rizzi2022prediction,rizzi2023modeling})}
        \label{fig:vehicle}
\end{figure}

We employ the noise models developed for a representative eVTOL aircraft configuration -- the NASA Revolutionary Vertical Lift Technology (RVLT) quadrotor reference vehicle, as shown in Figure~\ref{fig:nasavehicle}. Researchers at NASA Langley Research Center (LaRC) have utilized the Aeroacoustic ROtor Noise (AARON) tool within Aircraft NOise Prediction Program (ANOPP2) to generate noise hemispheres for this reference aircraft configuration, illustrated in Figure~\ref{fig:noisesphere}. This noise model is used in flight simulations to produce Sound Exposure Level (SEL) NPD data for three common operational modes: level flyover (L), dynamic departure (D), and approach (A).

Using the NPD data for the NASA RVLT quadrotor vehicle, we apply logarithmic interpolation to estimate noise levels at distances that were not directly measured. We fit regression models to the NPD data for all conditions. For each condition, defined by a specific combination of operational mode and measurement position, the regression model for single-event noise is expressed as follows:

\begin{equation}\label{eq:Rn}
    N_{\text{Single}} = c_0 + c_1 \log_{10}{z} + c_2 (\log_{10}{z})^2
\end{equation}
where $N_{\text{Single}}$ is noise in A-weighted SEL and $z \in [200, 20000]$ is distance in ft. Under different conditions, Table~\ref{tbl:fittednpds} includes the regression coefficients in Equation~\ref{eq:Rn}.

\begin{table}[htbp]
    \centering
    \caption{Regression coefficients for the NPD models}
    \begin{tabular}{llll}
    \hline
    Condition       & $c_0$    & $c_1$    & $c_2$    \\ \hline
    Mode L - Centerline & 88.09 & 3.21  & -2.62 \\ 
    Mode L - Side   & 78.01 & 7.26  & -3.39 \\ 
    Mode D - Centerline & 84.05 & 8.76  & -4.18 \\ 
    Mode D - Side   & 77.34 & 11.34 & -4.72 \\ 
    Mode A - Centerline & 93.35 & 5.17  & -2.86 \\ 
    Mode A - Side   & 85.55 & 6.83  & -3.14 \\ \hline
    \end{tabular}
    \label{tbl:fittednpds}
\end{table}

In this study, we focus on simulating the noise impact of UAM operations during flyover by primarily using the regression coefficients from the condition ``Mode L - Centerline'' located in the first row of Table~\ref{tbl:fittednpds}. In addition, we use the increase in noise above the local community's ambient level as a metric to assess UAM's community noise impact~\cite{gao2024noise}, as it correlates with perceived annoyance among residents. The cumulative noise increase over the ambient noise level $N_{\text{ambient}}$ of a given area on the ground is defined as:
\begin{equation}
    \begin{aligned}
        N_{\text{increase}} &= N_{\text{cumulative}} - N_{\text{ambient}}\\
        &= 10\log_{10}( \sum_{i = 1}^{N}10^{N_{\text{Single}}^i /10}) -  35.56  - N_{\text{ambient}}
    \label{eq:cumulative}
    \end{aligned}
\end{equation}

\subsection{Aircraft Energy Consumption Model}

The aircraft energy consumption model is another key component of aircraft performance in this study. In the proposed framework, we aim to jointly manage the noise impact and safety assurance of UAM through altitude adjustments of aerial vehicles. Specifically, within a UAM network, we define five cruising altitudes: 1,000 ft, 1,500 ft, 2,000 ft, 2,500 ft, and 3,000 ft. While the ability to adjust altitude may reduce the aircraft's noise impact by allowing for higher flight levels and ensuring separation assurance through the distribution of air traffic across various altitudes, this arrangement also has the downside of increased energy consumption. Therefore, energy consumption is a crucial factor in design tradeoffs.

In this work, we employ aircraft performance models to investigate how increasing altitude affects the energy consumption of eVTOL aircraft in UAM missions. A comprehensive analysis process is provided in Section~\ref{sec:Appendix} of the Appendix. The analysis follows the standard flight profile of eVTOL aircraft, comprising five segments: vertical takeoff, climb, cruise, descent, and vertical landing. For each segment, we determine the appropriate aircraft performance models to represent the power and energy consumption of a tiltrotor eVTOL aircraft configuration. By simulating flights at various altitudes throughout the Austin UAM network, we find that each increase of 500 ft in altitude leads to an average increase of 5\% in energy consumption for the entire trip, as shown in Figure~\ref{fig:EnergyAltitude} in the Appendix. This evidence will be utilized in training the reinforcement learning model and exploring tradeoffs associated with energy consumption. Interested readers can refer to Section~\ref{sec:Appendix} for further details.

\subsection{Reinforcement Learning Model}
Tactical separation assurance in UAM involves making decentralized decisions in uncertain and dynamic environments. DRL offers a natural solution by enabling aircraft to learn conflict-avoidance strategies directly from simulation, without requiring complete knowledge of the environment or explicit coordination between agents. Although multiple aircraft interact in the environment, the learning problem is formulated from the perspective of an individual aircraft. Each aircraft acts as an agent that observes its own state together with nearby neighboring aircraft, while the behavior of other agents is reflected through the resulting state transitions and rewards. Because all aircraft are homogeneous and share the same objective structure, a single policy is shared across agents during training and execution. Our learning framework is similar in spirit to PPO-based multi-agent reinforcement learning approaches such as MAPPO, but differs in that both the actor and critic operate on local observations and neighboring aircraft information is aggregated using the attention-based D2MAV-A architecture. Prior DRL approaches have shown success using local observations and advanced neural architectures \cite{brittain2020deep, DENIZ2024100157}, but most focus solely on maintaining lateral or vertical safe separation. In contrast, we develop a DRL framework that not only enforces vertical separation but also incorporates noise mitigation and energy consumption as additional objectives. Our goal is to understand the tradeoffs between these competing factors, enabling aircraft to make altitude adjustments that jointly balance safety, environmental impact, and energy efficiency in large-scale UAM operations.

\subsubsection{Key Reinforcement Learning Concepts}

Reinforcement learning (RL) provides a framework for training agents to make sequential decisions through repeated interaction with a dynamic environment~\cite{sutton2018reinforcement}. At each discrete timestep \( t \), the agent observes the current state \( s_t \), selects an action \( a_t \), and receives a scalar reward \( r_t \) that reflects the quality of the action in the context of the environment. This process produces a trajectory of states, actions, and rewards over time, which the agent uses to learn a policy that maps states to actions in a way that maximizes long-term performance.

Formally, we model the problem as a Markov decision process (MDP), defined by a tuple \( (S, A, T, R, \gamma) \), where \( S \) is the set of states, \( A \) is the set of actions, \( T(s_{t+1} \mid s_t, a_t) \) defines the probability of transitioning to state \( s_{t+1} \) given action \( a_t \) in state \( s_t \), \( R(s_t, a_t) \) defines the reward received after taking action \( a_t \) in state \( s_t \), and \( \gamma \in [0,1] \) is the discount factor that controls the weight of future rewards~\cite{sutton2018reinforcement}. In our setting, this MDP is defined from the perspective of a single aircraft operating in a shared multi-agent environment. The state used by the policy consists of the aircraft's own state together with information about nearby neighboring aircraft. Thus, while multiple aircraft make decisions simultaneously, each policy evaluation and update is performed using per-agent state, action, reward, and transition tuples. The agent's objective is to learn a policy $\pi: S \rightarrow \mathcal{P}(A)$ that maximizes the expected cumulative reward:

\begin{equation}
    J(\pi) = \mathbb{E}_{\pi} \left[\sum_{t=0}^{T} \gamma^t R(s_t, a_t)\right],
\end{equation}
where the expectation is taken over the trajectories induced by the policy $\pi$.

\paragraph{Value Functions} RL algorithms often make use of value functions to evaluate how good it is to be in a particular state or to take a certain action. The state-value function $V^\pi(s)$ is defined as the expected cumulative reward when starting in state $s$ and following policy $\pi$:

\begin{equation}
    V^\pi(s) = \mathbb{E}_\pi \left[ \sum_{t=0}^{T} \gamma^t R(s_t, a_t) \mid s_0 = s \right],
\end{equation}

The action-value function $Q^\pi(s, a)$ extends this idea to state-action pairs:

\begin{equation}
    Q^\pi(s, a) = \mathbb{E}_\pi \left[ \sum_{t=0}^{T} \gamma^t R(s_t, a_t) \mid s_0 = s, a_0 = a \right].
\end{equation}

\paragraph{Actor-Critic Methods} Many modern RL algorithms, including the one used in this work, are based on the actor-critic framework~\cite{konda1999actor}. This structure consists of two components:

\begin{itemize}
    \item \textbf{Actor:} The actor represents the policy $\pi_\theta(a \mid s)$, which selects actions based on the current state. The actor is typically implemented as a neural network parameterized by $\theta$.
    \item \textbf{Critic:} The critic estimates the value function $V_\phi^\pi(s)$ or $Q_\phi^\pi(s, a)$ and provides feedback to the actor about the quality of its actions. The critic is usually implemented as a neural network with separate parameters $\phi$.
\end{itemize}
During training, the critic reduces the variance in policy gradient estimates, enabling more stable learning. The actor is updated to favor actions that the critic evaluates as high-value, and the critic is updated to better approximate the true value function.

\subsubsection{Proximal Policy Optimization}

To train our actor-critic model, we use proximal policy optimization (PPO)~\cite{schulman2017proximal}, a first-order policy gradient method that has demonstrated strong empirical performance across a wide range of reinforcement learning tasks. PPO is designed to improve learning stability by constraining the amount that the policy can change at each update step.

\paragraph{Policy Gradient Objective}  
Policy gradient methods aim to directly optimize the expected return $J(\pi_\theta)$ by adjusting the policy parameters $\theta$ in the direction of the performance gradient:

\begin{equation}
    \nabla_\theta J(\pi_\theta) \approx \mathbb{E}_{t} \left[ \nabla_\theta \log \pi_\theta(a_t \mid s_t) A_t \right],
\end{equation}
where \(A_t\) is the advantage function, which estimates how much better taking action \(a_t\) in state \(s_t\) is compared to the average action. This advantage can be estimated using the value function:

\begin{equation}
    A_t = Q^\pi_\phi(s_t, a_t) - V^\pi_\phi(s_t).
\end{equation}

\paragraph{Clipped Surrogate Objective}  
Instead of applying large, unconstrained policy updates that may degrade performance, PPO introduces a \textit{clipped} surrogate objective to restrict how far the new policy $\pi_\theta$ can deviate from the old policy $\pi_{\theta_{\text{old}}}$:
\begin{equation}
    L^{\text{CLIP}}(\theta) = \mathbb{E}_t \left[
        \min \left(
            \psi_t(\theta) A_t,
            \text{clip}(\psi_t(\theta), 1 - \epsilon, 1 + \epsilon) A_t
        \right)
    \right],
\end{equation}
where \(\psi_t(\theta) = \pi_\theta(a_t \mid s_t)/\pi_{\theta_{\text{old}}}(a_t \mid s_t)\) is the probability ratio between the new and old policies. The clip function prevents \(\psi_t(\theta)\) from moving too far away from 1, discouraging overly large policy updates. This objective encourages improvement when the new policy outperforms the old policy (\(A_t > 0\)), but limits the size of updates to ensure stable learning. The hyperparameter \(\epsilon\) controls the strength of this constraint.

\paragraph{PPO Loss}  
PPO is trained with a total loss function that combines three components:
\begin{equation}
    L^{\text{PPO}}(\theta, \phi) = L^{\text{CLIP}}(\theta) + \lambda_1 \cdot L^{\text{VF}}(\phi) - \lambda_2 \cdot H(\pi_\theta),
\end{equation}
where \(L^{\text{CLIP}}(\theta)\) is the clipped surrogate objective that encourages policy improvement while avoiding overly large updates, \(L^{\text{VF}}(\phi)\) is the Huber loss between the predicted and target value functions used to train the critic, and \(H(\pi_\theta)\) is the entropy of the policy, which encourages exploration by penalizing deterministic action distributions. The coefficients \(\lambda_1\) and \(\lambda_2\) control the relative importance of the value function loss and policy entropy, respectively.

\begin{figure}
    \centering
    \includegraphics[width=1.0\linewidth]{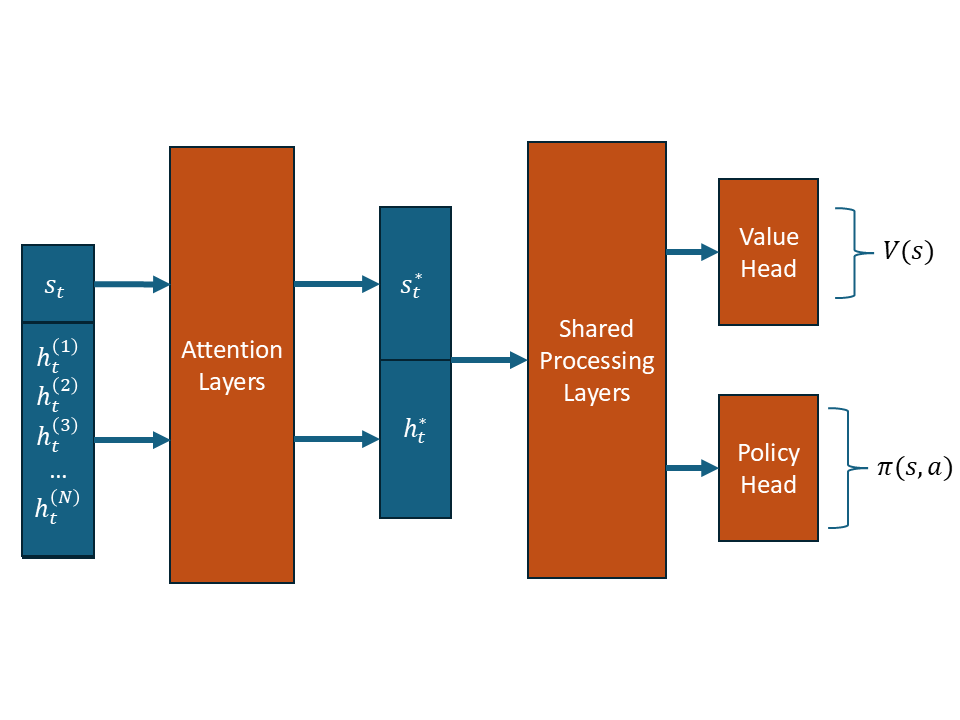}
    \caption{Architecture of the D2MAV-A network used in our experiments, adapted from \cite{brittain2020deep}. The network processes ownship and neighboring aircraft states, uses attention-based aggregation to compute a fixed-size representation, and outputs both a policy and a value function for the aircraft.}
    \label{fig:d2mav-a}
\end{figure}

\subsubsection{Network Architecture and D2MAV-A Design}
To implement the PPO algorithm for vertical separation and noise-aware planning, we adopt the D2MAV-A architecture \cite{brittain2020deep}, a deep actor-critic model tailored for decentralized air traffic management. The architecture is specifically designed to operate under partial observability and accommodate a variable number of neighboring aircraft by leveraging an attention-based aggregation mechanism. A visualization of the D2MAV-A architecture is provided in \cref{fig:d2mav-a}. Figure~\ref{fig:d2mav-a} outlines the full processing pipeline from raw inputs to policy and value outputs.

\paragraph{Input Representation} 
At each timestep \( t \), the model receives as input both the state of the ownship aircraft and the states of all observed neighboring aircraft within a fixed communication radius:
\begin{itemize}
    \item The \textbf{ownship state} \( s_t \) includes the aircraft's current state information.
    \item Each \textbf{neighbor state} \( h_t^{(i)} \) includes state information on the \( i \)-th neighboring aircraft from the perspective of the ownship.
\end{itemize}
These input components are depicted on the left side of \cref{fig:d2mav-a}, where separate input blocks are used for the ownship and each neighboring aircraft.

\paragraph{Feature Encoding and Attention-Based Aggregation}
To process variable-length sets of neighbors, D2MAV-A uses an attention mechanism that produces a fixed-size summary of the neighborhood. The ownship state \( s_t \) is first embedded as a fixed-length vector \( s_t^* \) using a fully connected layer. Similarly, each neighbor state \( h_t^{(i)} \) is passed through a shared fully connected layer to produce the encoded neighbor representation \( \bar{h}_t^{(i)} \).

The model then computes attention weights over the neighbors using a bilinear compatibility function between the ownship embedding and each neighbor embedding:
\begin{equation}
    \alpha_i = \frac{\exp\left({s_t^*}^\top W_1 \bar{h}_t^{(i)}\right)}{\sum_j \exp\left({s_t^*}^\top W_1 \bar{h}_t^{(j)}\right)},
\end{equation}
where \( W_1 \) is a learned weight matrix. These weights are used to compute an attention-weighted context vector:
\begin{equation}
    c_t = \sum_i \alpha_i \bar{h}_t^{(i)},
\end{equation}
which is then projected through a learned transformation:
\begin{equation}
    h_t^* = \tanh(W_2 c_t),
\end{equation}
where \( W_2 \) is another learned weight matrix. This attention-based aggregation mechanism is shown in the center of \cref{fig:d2mav-a}, highlighting the transformation from variable-length neighbor inputs to a single context vector.

\paragraph{Policy and Value Estimation}
The model concatenates the ownship embedding \( s_t^* \) with the aggregated neighbor representation \( h_t^* \) and passes the result through shared fully connected layers. This intermediate representation is then routed to two task-specific output heads:
\begin{itemize}
    \item The \textbf{policy head} outputs action probabilities over the discrete set of actions \( A \) using a softmax activation.
    \item The \textbf{value head} estimates the state value \( V(s_t) \) for use in the PPO update.
\end{itemize}
The final output heads corresponding to the policy and value estimates are depicted on the right side of \cref{fig:d2mav-a}, completing the forward pass of the network.

\paragraph{Centralized Training and Decentralized Execution}
Training is conducted in simulation using centralized data collection across multiple agents. Experience tuples \( (s_t, a_t, r_t, s_{t+1}) \) collected from all aircraft are aggregated into shared training batches and used to optimize a single shared policy using PPO. After training, identical copies of the policy network are deployed to all agents. During execution, each agent selects actions based solely on its own local state and the observed states of nearby aircraft, enabling fully decentralized operation.

\begin{figure}
    \centering
    \includegraphics[width=1.1\linewidth]{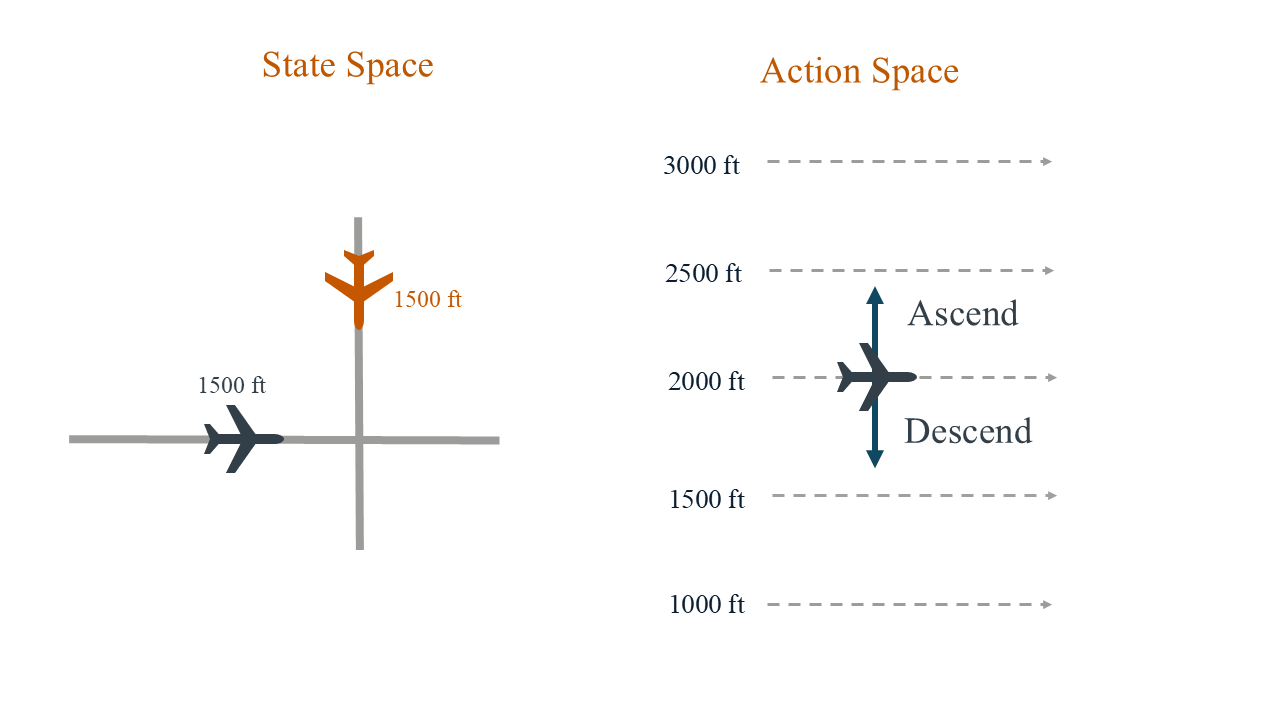}
    \caption{Illustration of the agent’s state and action spaces. The state includes the positions and altitudes of the ownship aircraft as well as the positions and altitudes of neighboring aircraft (left), while the action space (right) consists of discrete vertical adjustments, allowing the agent to ascend, descend, or maintain its current altitude within a fixed set of flight levels.}
    \label{fig:state-action-space}
\end{figure}

\subsection{UAM Network Operations as a Markov Decision Process}

We model the UAM environment as a multi-agent system in which multiple aircraft operate simultaneously and interact through shared airspace constraints. For decision-making and learning, the problem is formulated from the perspective of an individual aircraft as a Markov decision process (MDP). In this formulation, each aircraft observes its own state and local information about nearby aircraft, selects a discrete altitude-adjustment action, and receives a reward based on the resulting safety, noise, and energy outcomes. The MDP is implemented within the BlueSky air traffic simulator \cite{Simulator}, which governs the low-level dynamics and state transitions.

\subsubsection{State and Action Space}

Each aircraft observes an ownship state vector that encodes features relevant to decision-making:
\begin{equation}
    s_{t} = \{z, b_{\text{changing}}, z_{\text{target}}, a_{t-1}, \mathcal{N}_{\text{adj}}\},
\end{equation}
where \(z\) denotes the current altitude, \(b_{\text{changing}}\) indicates whether an altitude change is currently in progress, \(z_{\text{target}}\) is the goal altitude associated with the current maneuver, \(a_{t-1}\) is the previous action taken by the aircraft, and \(\mathcal{N}_{\text{adj}}\) is the number of altitude increases the aircraft has made so far. 
The current altitude provides information relevant to both noise exposure and vertical separation, while \(b_{\text{changing}}\), \(z_{\text{target}}\), and \(a_{t-1}\) capture the context of ongoing altitude maneuvers that persist across multiple decision steps. 
Including these variables allows the policy to distinguish between otherwise identical instantaneous altitudes that correspond to different maneuvering states, thereby preserving the Markov property of the decision process. 
Finally, \(\mathcal{N}_{\text{adj}}\) captures the cumulative number of altitude adjustments and is used to model energy costs associated with repeated altitude changes.

To ensure vertical separation, aircraft also receive information from nearby neighboring aircraft within a communication range. A neighboring aircraft is defined as any aircraft whose horizontal distance from the ownship is less than a fixed communication radius \(d_{\text{comm}}\). All aircraft satisfying this condition are included in the neighbor set at each decision step and are ordered according to aircraft ID to provide a consistent input structure. The observed state for a neighbor \(i\) is given by:
\begin{equation}
    h_t^{(i)} = \{z_{\text{rel}}^{(i)}, d_o^{(i)}, a_{t-1}^{(i)}\},
\end{equation}
where \(z_{\text{rel}}^{(i)}\) is the relative altitude, \(d_o^{(i)}\) is the horizontal distance between the ownship and the neighbor, and \(a_{t-1}^{(i)}\) is the previous action of the neighbor. 
These variables provide the information required to assess potential separation conflicts and anticipate the short-term motion of neighboring aircraft.

The action space is defined over altitude adjustments:
\begin{equation}
    A = \{v_{z-}, v_{z0}, v_{z+}\},
\end{equation}
where \(v_{z-}\) and \(v_{z+}\) represent descending and ascending, respectively, and \(v_{z0}\) denotes level flight. 
To ensure commitment to altitude changes, once an action is initiated, the aircraft must reach \(z_{\text{target}}\) before making further adjustments. 
This temporal coupling enforces action consistency and prevents oscillatory behavior during training.

\subsubsection{Reward Function}

At each timestep, every aircraft receives its own local reward computed from its current state, observed neighboring aircraft, and chosen action. All agents use the same reward structure, so the setting is cooperative in the sense that agents optimize aligned objectives, even though no explicit team-level global reward is used. The agent's reward signal includes and balances three potentially competing objectives: minimizing noise impact, ensuring vertical separation, and conserving energy.

\paragraph{Noise Reward}  
Based on the SEL model in Equation~\ref{eq:Rn}, the noise reward is defined as:
\begin{equation} \label{eq:Rs}
    R_{\text{noise}}(s_{t}) = \frac{N_{\text{Single}} - N_{\text{min}}}{N_{\text{max}} - N_{\text{min}}},
\end{equation}
where \(N_{\text{Single}}\) is the current aircraft's noise contribution, normalized against the maximum and minimum possible noise levels $N_{\max}$ and $N_{\min}$ at \(z_{\text{min}}\) and \(z_{\text{max}}\), respectively.

\paragraph{Separation Reward}  
To penalize altitude-level congestion more precisely, we use a \textit{proximity-weighted congestion penalty}. For each aircraft \(i\), we consider nearby aircraft flying within a narrow vertical band (i.e., within a loss of separation distance) and within communication distance. Let \( d_o^{(j)} \) denote the horizontal distance from aircraft \(i\) to neighbor \(j\), and let \( d_{\text{LOS}} \) and \( d_{\text{comm}} \) be the minimum separation threshold and the maximum communication threshold, respectively. The proximity weight for each neighbor is defined as
\begin{equation}
w_j = 
\begin{cases}
1 & \text{if } d_o^{(j)} < d_{\text{LOS}}, \\
\displaystyle \frac{d_{\text{comm}} - d_o^{(j)}}{d_{\text{comm}} - d_{\text{LOS}}} & \text{if } d_{\text{LOS}} \leq d_o^{(j)} \leq d_{\text{comm}}, \\
0 & \text{otherwise}.
\end{cases}
\end{equation}

We define the set of neighbors considered to be at the same altitude as
\begin{equation}
\mathcal{N}_{\text{same}}^{(i)} = \left\{ j \mid |z_{\text{rel}}^{(j)}| < d_{\text{LOS}},\ j \neq i \right\},
\end{equation}
where \( z_{\text{rel}}^{(j)} \) is the vertical offset of aircraft \(j\) relative to aircraft \(i\), and \( d_{\text{LOS}} \) is the separation threshold.
The total weighted congestion for aircraft \(i\) is computed by summing proximity weights over this set
$C_i = \sum_{j \in \mathcal{N}_{\text{same}}^{(i)}} w_j,$
and the corresponding separation penalty is:
\begin{equation}
R_{\text{separation}}(s_t, h_t) = - \min\left( \frac{C_i}{C_{\text{max}}}, 1 \right),
\end{equation}
where \( C_{\text{max}} = 10 \) is the maximum expected weighted congestion. This formulation penalizes both the number and proximity of neighboring aircraft at the same altitude, encouraging vertical separation.

\paragraph{Energy Reward}
To discourage excessive altitude changes, we introduce an energy penalty that scales with the number of upward adjustments previously taken by each aircraft. At each timestep, we track whether the agent performs an ascent. If an aircraft climbs for 500 ft to a higher flight level, the aircraft's adjustment counter $\mathcal{N}_{\text{adj}}$ is incremented and used to scale the energy penalty:

\begin{equation}
    R_{\text{energy}}(s_t, a_t) = 
    \begin{cases}
        -c_e \cdot \mathcal{N}_{\text{adj}} & \text{if } a_t = v_{z+}, \\
        0 & \text{otherwise},
    \end{cases}
\end{equation}
where $c_e = 0.05$ is a fixed energy cost factor. This formulation penalizes each additional ascent more heavily than the last, encouraging aircraft to limit the number of altitude adjustments over the course of a scenario. Descents and level flight actions do not incur any energy penalty, as supported by the analysis in Section~\ref{sec:Appendix}.

\paragraph{Total Reward}  
The per-agent reward function is a weighted sum of the three objectives:
\begin{equation} \label{eq:Rf}
\begin{aligned}
    &R(s_t, h_t, a_t)= \\
    & \rho_{\text{noise}} R_{\text{noise}}(s_t) + \rho_{\text{sep}} R_{\text{separation}}(s_t, h_t) + \rho_{\text{energy}} R_{\text{energy}}(s_t, a_t),
\end{aligned}
\end{equation}
where the coefficients \(\rho_{\text{noise}}\), \(\rho_{\text{sep}}\), and \(\rho_{\text{energy}}\) control the tradeoff between the three objectives -- noise mitigation, safe separation, and energy consumption.

\begin{table*}[h!]
\centering
\caption{Training hyperparameters and environment constants for D2MAV-A PPO}
\label{tab:ppo_hyperparams}
\begin{tabular}{|l|c|p{7.5cm}|}
\hline
\textbf{Parameter} & \textbf{Value} & \textbf{Description} \\
\hline
\multicolumn{3}{|c|}{\textit{PPO Training Hyperparameters}} \\
\hline
Batch size ($B$) & 512 & Number of transition tuples used per PPO update. \\
Number of epochs ($K$) & 6 & Number of gradient updates per PPO iteration. \\
Learning rate ($\eta$) & $1 \times 10^{-5}$ & Step size for updating actor and critic parameters. \\
Clipping parameter ($\epsilon$) & 0.4 & Maximum allowable change in policy probability ratio. \\
Value loss coefficient ($\lambda_1$) & $0.01$ & Weight of the value function loss in the PPO loss. \\
Entropy coefficient ($\lambda_2$) & $0.01$ & Weight of the entropy bonus in the PPO loss. \\
Discount factor ($\gamma$) & 0.99 & Weight assigned to future rewards in return estimation. \\
Neural network size & 256 nodes & Hidden layer size for both actor and critic networks. \\
PPO update interval & 32 steps & Number of simulation steps between PPO updates. \\
Number of PPO iterations & 10,000 & Total training updates performed on shared policy network. \\
Parallel simulations & 5 & Number of environment instances used to collect training batches. \\
\hline
\multicolumn{3}{|c|}{\textit{Simulation and Environment Parameters}} \\
\hline
Number of aircraft & 136 & Fixed number of aircraft in each training and evaluation scenario. \\
Communication distance ($d_{\text{comm}}$) & 2.5 km & Maximum range for detecting neighbors in local observations. \\
Loss of separation threshold ($d_{\text{LOS}}$) & 150 m & Minimum allowed Euclidean distance between aircraft. \\
Congestion normalization ($C_{\text{max}}$) & 10 & Normalization constant in separation reward penalty. \\
Energy cost factor ($c_e$) & 0.05 & Penalty scaling factor for energy consumed during ascents. \\
Altitude levels & 5 (1000–3000 ft) & Discrete altitudes for lateral flight, spaced by 500 ft. \\
Noise normalization ($N_{\text{min}}$) & 67.54 dB SEL & Minimum single-event noise level at 3000 ft, used for reward normalization. \\
Noise normalization ($N_{\text{max}}$) & 74.14 dB SEL & Maximum single-event noise level at 1000 ft, used for reward normalization. \\
\hline
\end{tabular}
\end{table*}

\section{Experiments}\label{sec:exp}

\subsection{Scenario Description and Simulation Platform}

\begin{figure}[h!]
    \centering
    \begin{subfigure}[b]{0.35\textwidth}
        \centering
        \includegraphics[width=\textwidth]{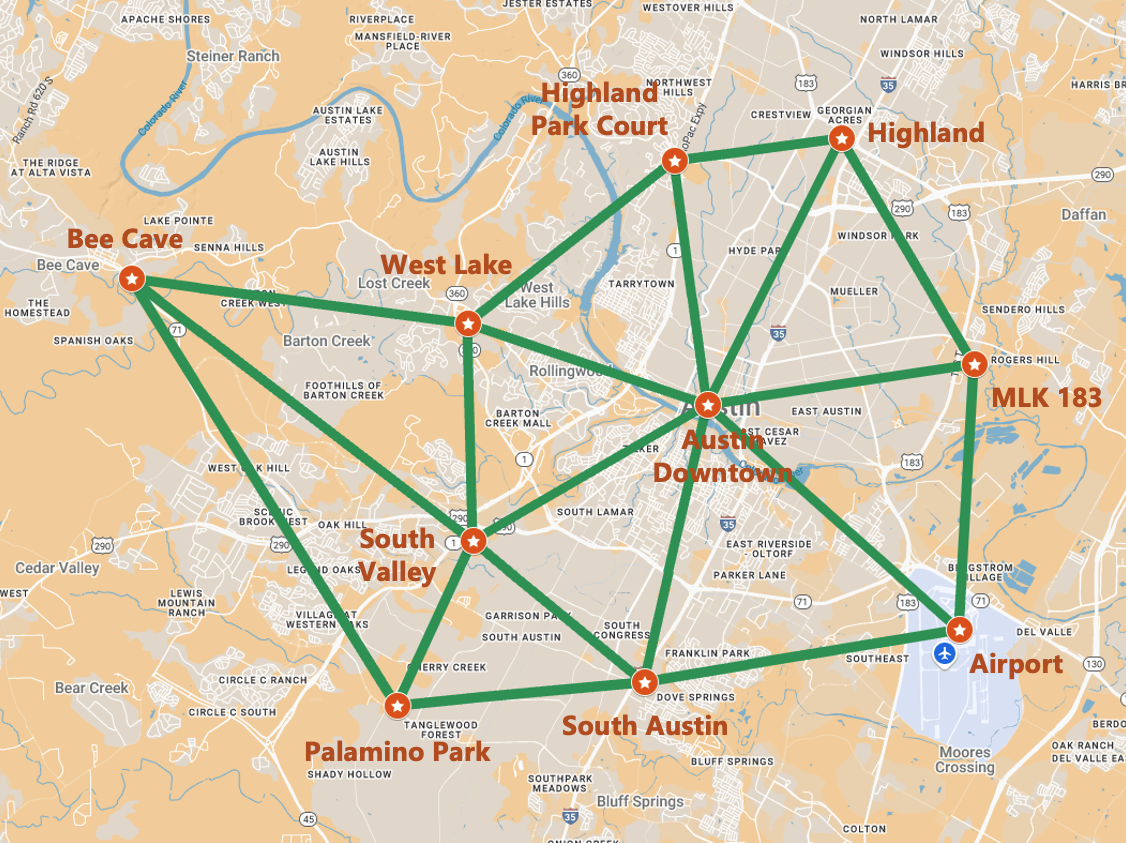}
        \caption{The South Austin UAM network}
        \label{fig:scenarioEnv}
    \end{subfigure}
    \hfill
    \begin{subfigure}[b]{0.39\textwidth}
        \centering
        \includegraphics[width=\textwidth,height=0.6\textwidth]{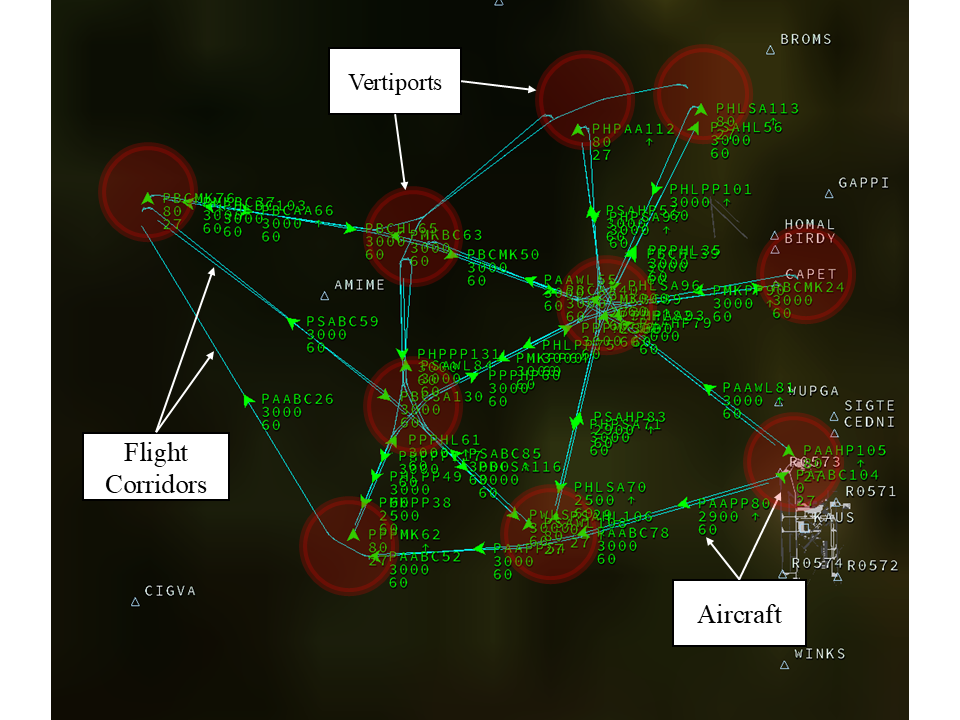}
        \caption{The South Austin network in the BlueSky simulator}
        \label{fig:scenarioEnv_sim}
    \end{subfigure}
    \caption{Overview of the South Austin UAM scenario.}
    \label{fig:scenarioEnv_combined}
\end{figure}

We evaluate our approach in a simulated UAM environment modeled after the South Austin region. The network consists of 10 vertiports and 19 unidirectional flight corridors, yielding 38 directional links across five altitude layers. Each flight corridor supports altitudes from 1,000 feet to 3,000 feet in 500-foot intervals. A total of 28 origin-destination (O-D) pairs are defined within the network. The structure of the network is shown in \cref{fig:scenarioEnv}.

We implement and execute all simulations using the BlueSky Air Traffic Simulator~\cite{Simulator}, an open-source platform designed for air traffic management research. Each scenario is defined using a text-based file that specifies the routes and takeoff times for all aircraft in the network. BlueSky provides real-time aircraft states at each simulation step and supports dynamic altitude changes through its command interface.

\subsection{Aircraft and Operational Assumptions}

The scenario includes 136 aircraft, each assigned a unique origin and destination at the start of the simulation. Aircraft follow fixed routes composed of pre-defined flight corridors and cannot switch to alternative routes once en route. Each aircraft operates under performance characteristics similar to those of the Eurocopter EC135.

Aircraft maintain awareness of nearby traffic within a horizontal communication range of \(d_{\text{comm}} = 2.5\,\text{km}\). A loss of separation (LOS) is defined as a pair of aircraft coming within \(d_{\text{LOS}} = 150\,\text{m}\) of each other, measured using Euclidean distance. Aircraft departing from the same origin are assumed to be initially separated by sufficient time and space to prevent LOS events. During flight, each aircraft monitors only those on intersecting flight corridors when evaluating potential conflicts. The reinforcement learning models are trained using repeated simulations of the same scenario, with aircraft taking off at the same time and on the same routes.

\subsection{Evaluation Metrics}

The aircraft aim to (i) avoid LOS events, (ii) minimize cumulative noise impact, and (iii) reduce energy consumption. These objectives are used to guide policy learning and are evaluated by testing the policies in $100$ episodes of the same scenario used in training. We report the following metrics during testing:
\begin{enumerate}
    \item Number of LOS events observed during the simulation.
    \item Cumulative noise increase over ambient levels, computed using Equation~\ref{eq:cumulative}.
    \item Number of altitude changes performed by all aircraft.
    \item Distribution of aircraft altitudes over the course of the simulation.
\end{enumerate}

The altitude distribution is computed by aggregating aircraft altitude values across all simulation timesteps. This provides a complete view of how frequently each altitude level is utilized, rather than a snapshot at selected moments.

To account for spatial variations in ambient noise, the environment is partitioned into zones, with each flight corridor and vertiport assigned an ambient noise level. These values are listed in the Appendix. Cumulative noise increase is computed by comparing the modeled UAM-generated noise in each zone against its ambient baseline. Lower cumulative noise increase indicates reduced impact on community noise exposure.

\begin{figure*}[h!]
    \centering
    \begin{subfigure}{0.48\linewidth}
        \centering
        \includegraphics[width=0.85\linewidth]{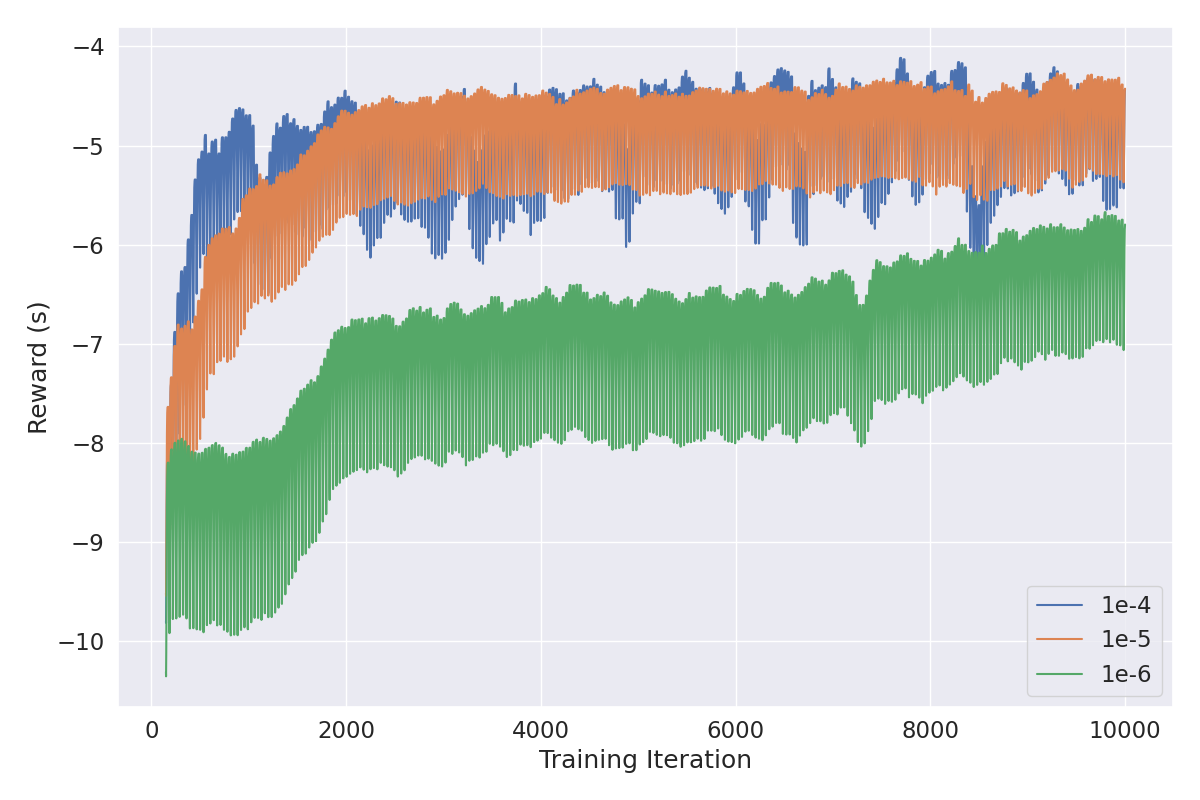}
        \caption{Learning Rate}
        \label{fig:hs_lr}
    \end{subfigure}
    \hfill
    \begin{subfigure}{0.48\linewidth}
        \centering
        \includegraphics[width=0.85\linewidth]{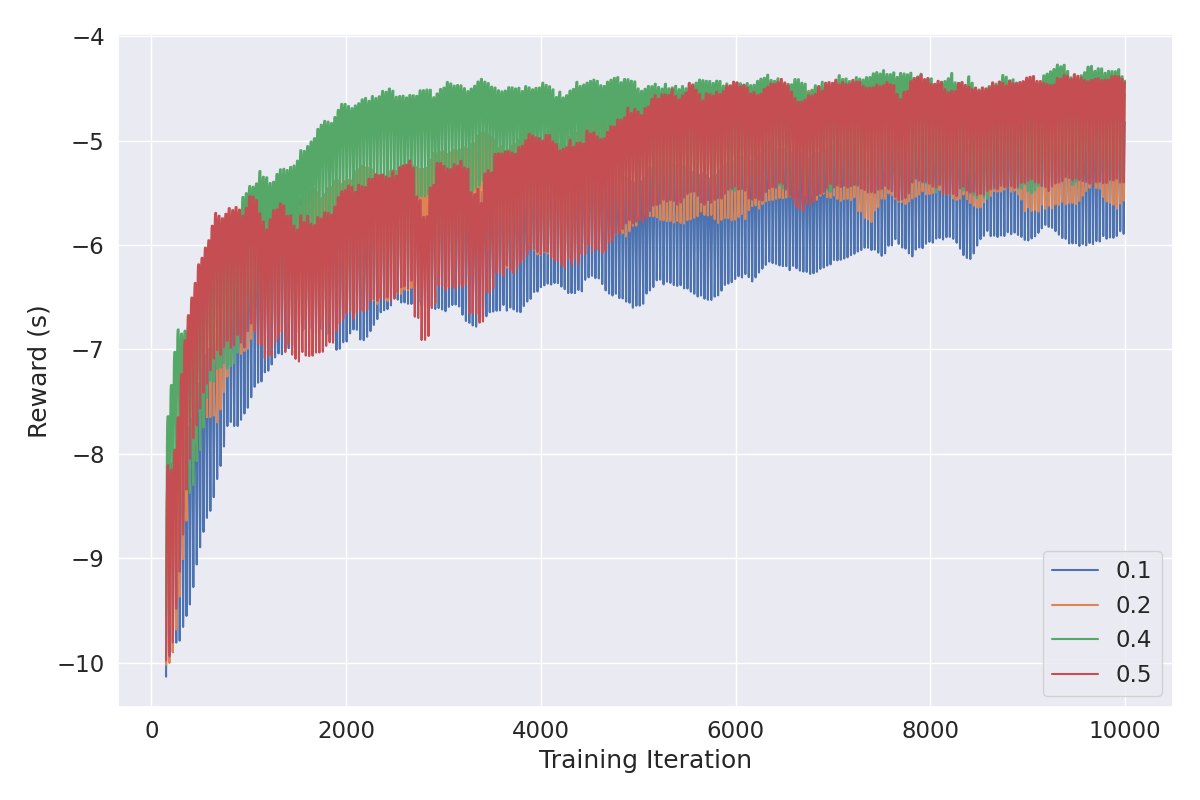}
        \caption{PPO Clipping Parameter}
        \label{fig:hs_ppo}
    \end{subfigure}
    \caption{Hyperparameter study for learning rate and PPO clipping parameter. Results show that the chosen values of $10^{-5}$ and $0.4$ lead to the best training performance.}
    \label{fig:hyperparameter-study}
\end{figure*}

\subsection{Hyperparameter Sensitivity Study}

To ensure stable and efficient training, we conducted a targeted hyperparameter sensitivity study on two key PPO parameters: the learning rate and the clipping parameter. These parameters strongly influence policy optimization dynamics and are known to affect both convergence speed and training stability. All other hyperparameters were held fixed during this study. For each candidate configuration, the policy was trained in the South Austin scenario described previously, and training performance was evaluated using the average episodic reward over training iterations. The resulting training curves are shown in \cref{fig:hyperparameter-study}.

\paragraph{Learning Rate}
We evaluated learning rates of $10^{-4}$, $10^{-5}$, and $10^{-6}$ to examine the tradeoff between convergence speed and training stability. When using a learning rate of $10^{-4}$, the training process exhibits significant oscillations in episodic reward, indicating unstable policy updates and difficulty converging to a consistent solution. At the opposite extreme, a learning rate of $10^{-6}$ produces stable training but results in slow policy improvement, requiring substantially more iterations to approach a comparable performance level. A learning rate of $10^{-5}$ provides the best balance between these two effects, enabling stable training while maintaining a reasonable convergence rate. Based on this observation, we adopt $10^{-5}$ as the learning rate for all experiments reported in this work.

\paragraph{Clipping Parameter}
We also evaluated several values of the PPO clipping parameter $\epsilon$, including $0.1$, $0.2$, $0.4$, and $0.5$. The clipping parameter controls the allowable magnitude of policy updates during training, thereby regulating the stability of policy improvement steps. Smaller values such as $0.1$ and $0.2$ constrain policy updates more aggressively, which leads to slower convergence and reduced exploration of improved policies. Conversely, larger values such as $0.5$ allow larger policy updates but introduce increased variability in training performance. Among the tested configurations, a clipping parameter of $0.4$ achieves the fastest and most consistent convergence, producing smoother reward growth compared to the other settings. Although this value is larger than commonly used PPO defaults, our empirical results indicate that it is well suited to the dynamics of the UAM conflict-resolution task. Therefore, we use $\epsilon = 0.4$ in all reported experiments.

\subsection{Pairwise Objective Tradeoffs}

We evaluate the effect of reward weighting on aircraft behavior by conducting three sets of experiments, each examining the tradeoff between a pair of objectives: (i) noise impact vs. vertical separation, (ii) energy consumption vs. vertical separation, and (iii) energy consumption vs. noise impact.

In each experiment, we fix the weight of one objective to zero and vary the remaining two reward coefficients such that their sum is equal to 1. Specifically, for a given pair of objectives \((\rho_a, \rho_b)\), we set \(\rho_a \in [0.0, 1.0]\), \(\rho_b = 1 - \rho_a\), and \(\rho_c = 0\), where \(c\) is the unused objective. Each resulting policy is trained for 10,000 iterations in the scenario described previously and evaluated over 100 independent testing episodes. Testing scenarios are held constant across all experiments but are separate from the training rollouts used for learning.

The goal of these experiments is to characterize how changes in reward composition influence key performance metrics, including cumulative noise increase, number of LOS events, number of altitude changes, and aircraft altitude distribution. Rather than benchmarking against external centralized or heuristic baselines, the focus is on understanding how decentralized policies behave as the relative importance of the objectives changes. This allows us to assess the behavioral and operational tradeoffs associated with prioritizing one objective over another.

\section{Results}\label{sec:results}

We evaluate the trained policies under the same scenario used during training, focusing on how reward weightings (\(\rho_\text{sep}, \rho_\text{noise}, \rho_\text{energy}\)) affect aircraft behavior and system-wide outcomes. Our analysis examines training performance, altitude selection, noise impact, and safety tradeoffs across 100 test episodes per policy
\footnote{Code for the simulation testing environment is provided at~\url{https://github.com/suryakmurthy/Integrated_Noise_Safety_UAM}}.

\begin{figure*}[t]
    \centering
    \begin{subfigure}[b]{0.3\textwidth}
        \centering
        \includegraphics[width=\textwidth]{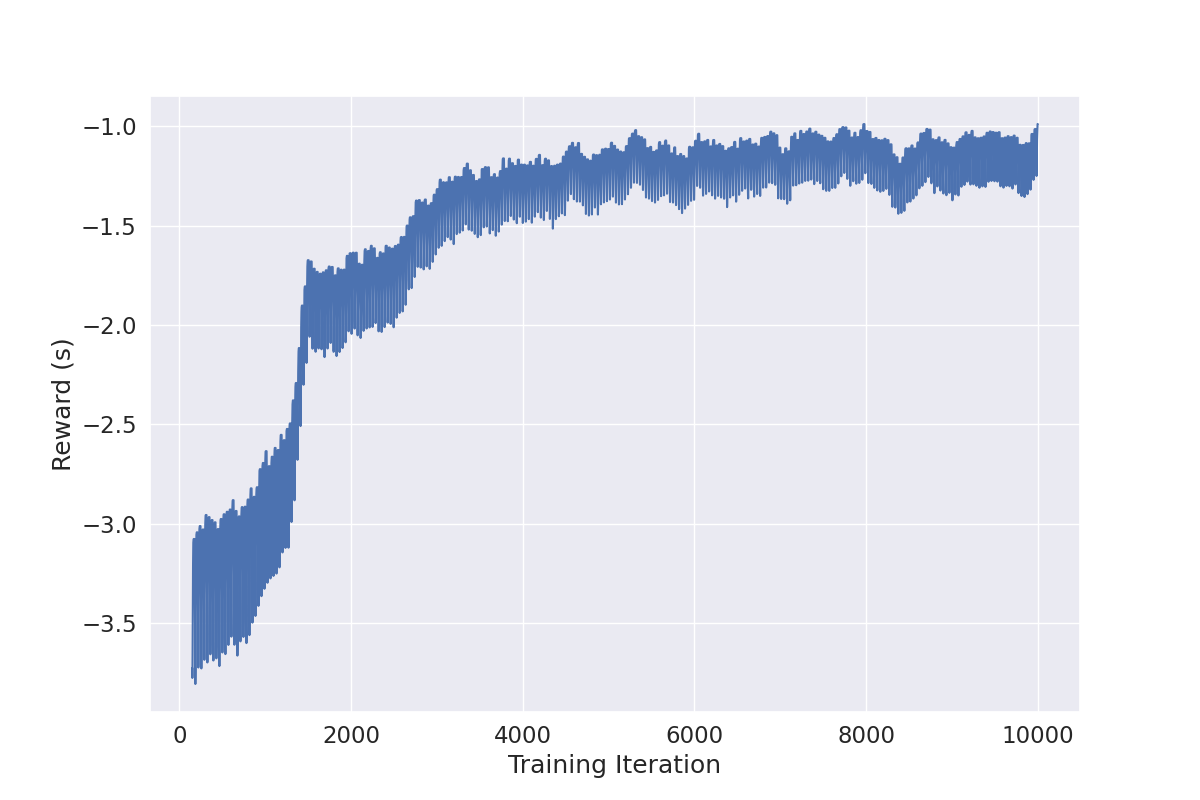}
        \caption{\(\rho_{\text{sep}}=1.0,\ \rho_{\text{noise}}=0,\ \rho_{\text{energy}}=0\)}
        \label{fig:training-separation-only}
    \end{subfigure}
    \hfill
    \begin{subfigure}[b]{0.3\textwidth}
        \centering
        \includegraphics[width=\textwidth]{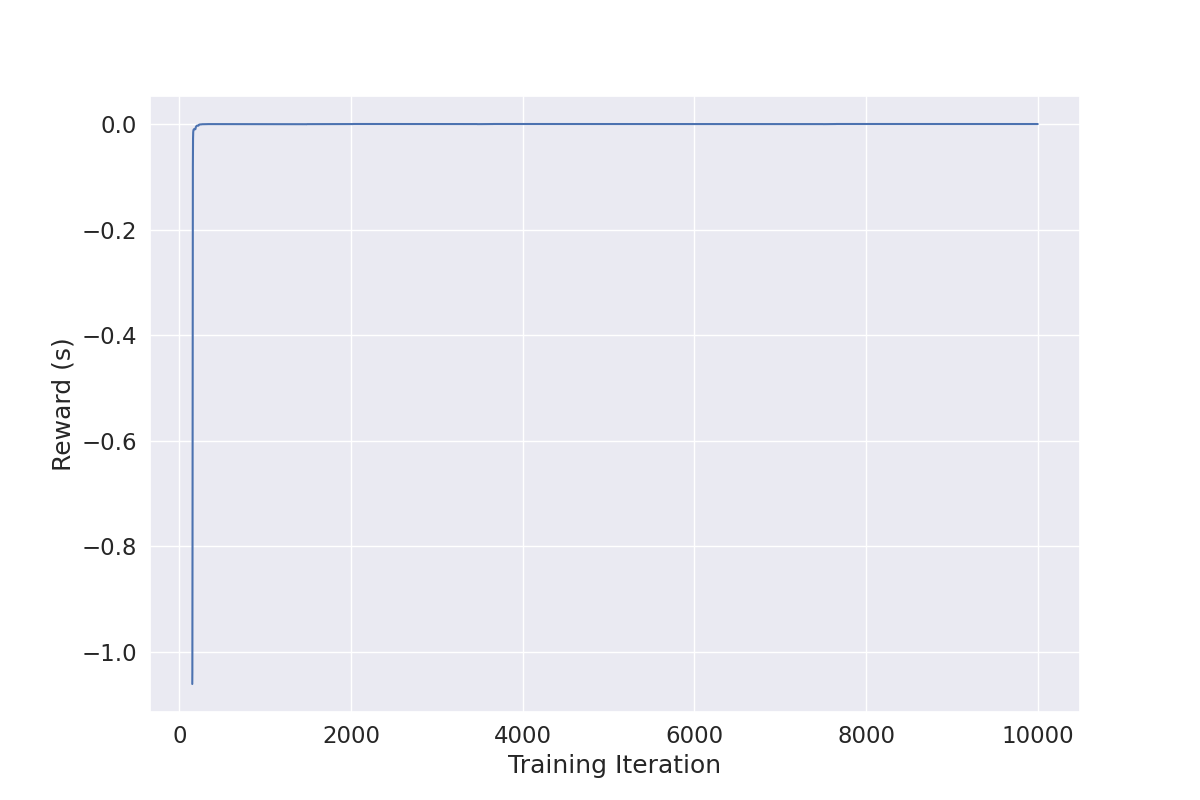}
        \caption{\(\rho_{\text{sep}}=0,\ \rho_{\text{noise}}=0,\ \rho_{\text{energy}}=1.0\)}
        \label{fig:training-energy-only}
    \end{subfigure}
    \hfill
    \begin{subfigure}[b]{0.3\textwidth}
        \centering
        \includegraphics[width=\textwidth]{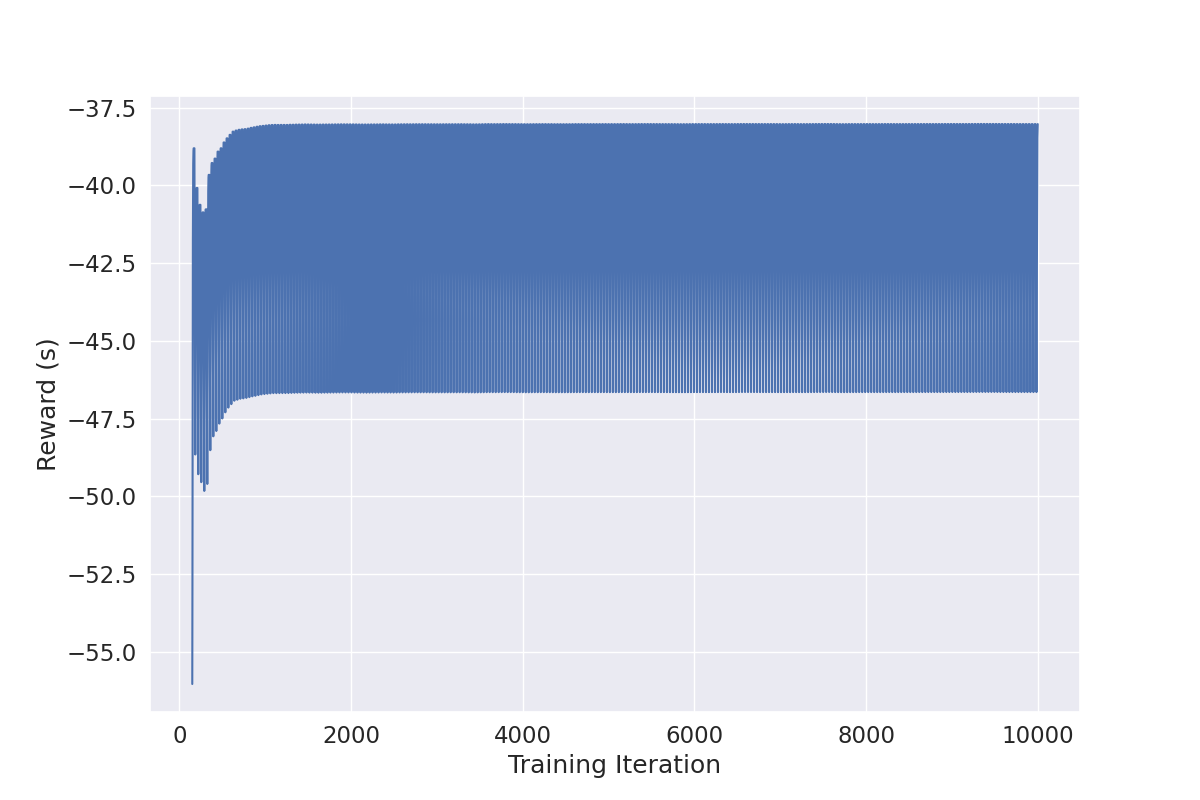}
        \caption{\(\rho_{\text{sep}}=0,\ \rho_{\text{noise}}=1.0,\ \rho_{\text{energy}}=0\)}
        \label{fig:training-noise-only}
    \end{subfigure}

    \vspace{1em}

    \begin{subfigure}[b]{0.3\textwidth}
        \centering
        \includegraphics[width=\textwidth]{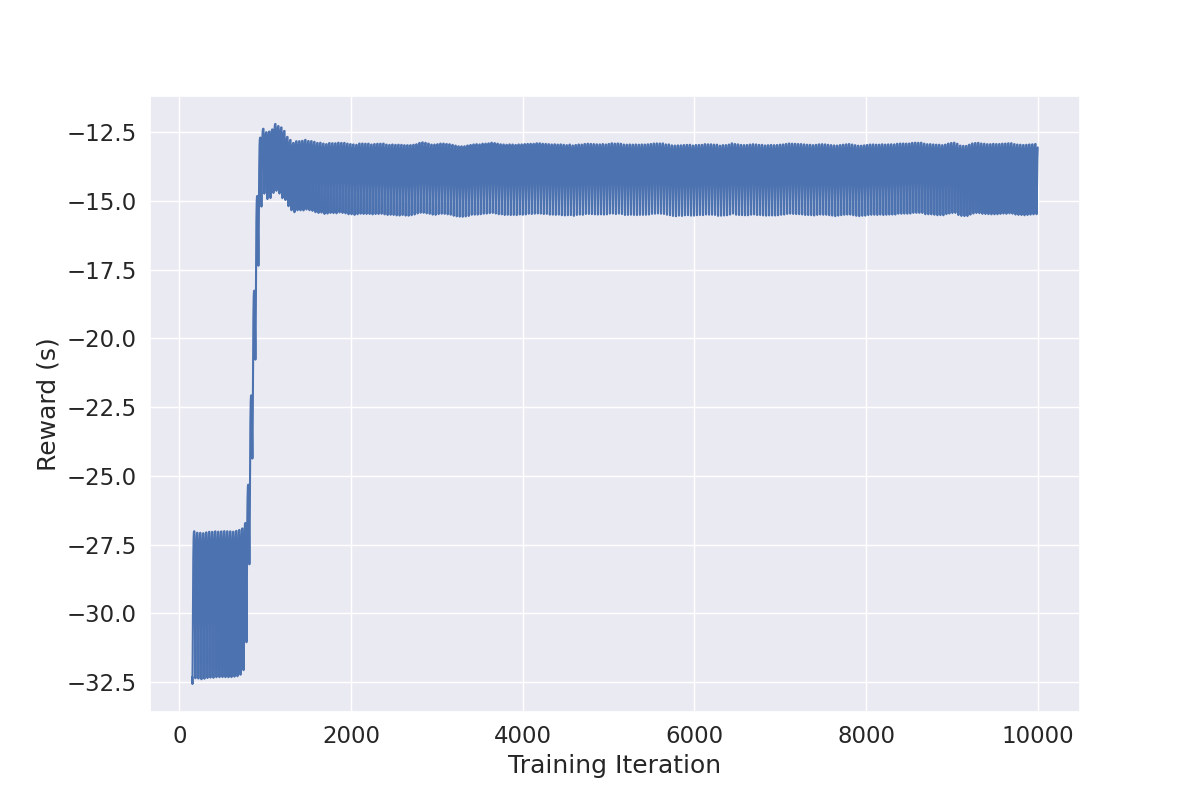}
        \caption{\(\rho_{\text{sep}}=0,\ \rho_{\text{noise}}=0.1,\ \rho_{\text{energy}}=0.9\)}
        \label{fig:training-noise-energy}
    \end{subfigure}
    \hfill
    \begin{subfigure}[b]{0.3\textwidth}
        \centering
        \includegraphics[width=\textwidth]{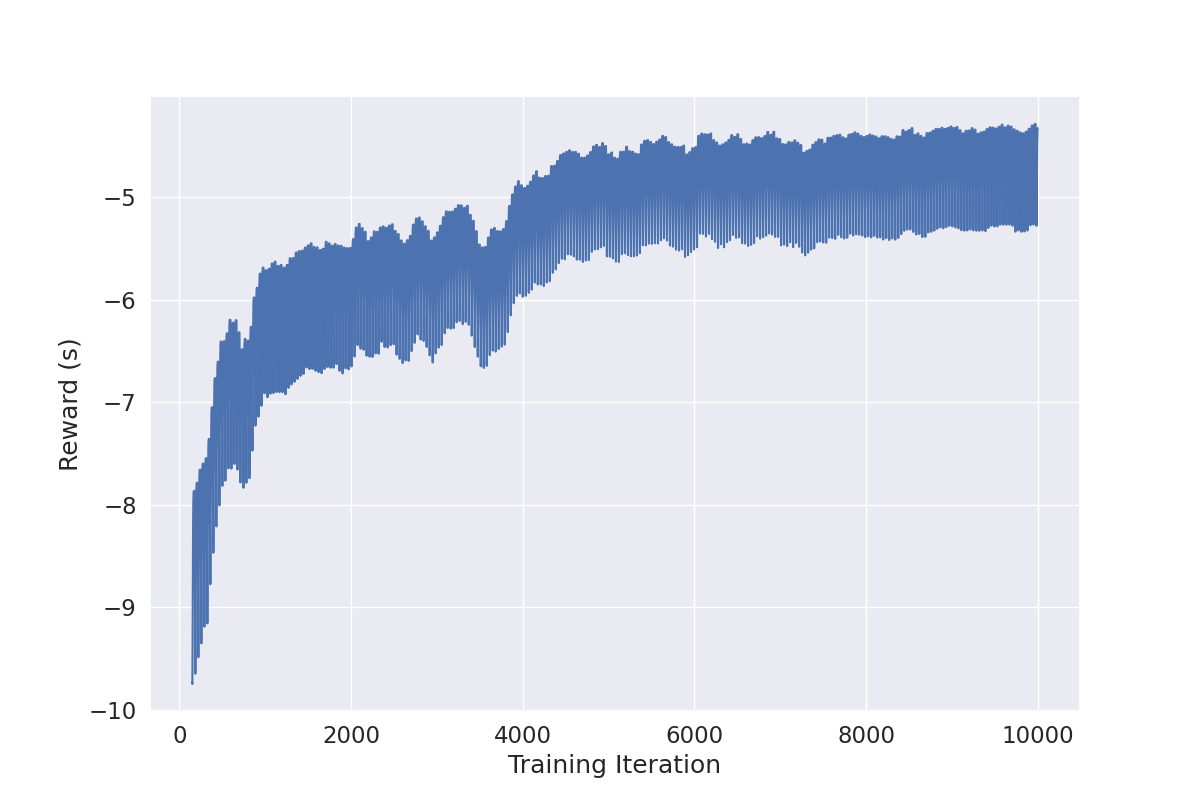}
        \caption{\(\rho_{\text{sep}}=0.95,\ \rho_{\text{noise}}=0.05,\ \rho_{\text{energy}}=0\)}
        \label{fig:training-noise-separation}
    \end{subfigure}
    \hfill
    \begin{subfigure}[b]{0.3\textwidth}
        \centering
        \includegraphics[width=\textwidth]{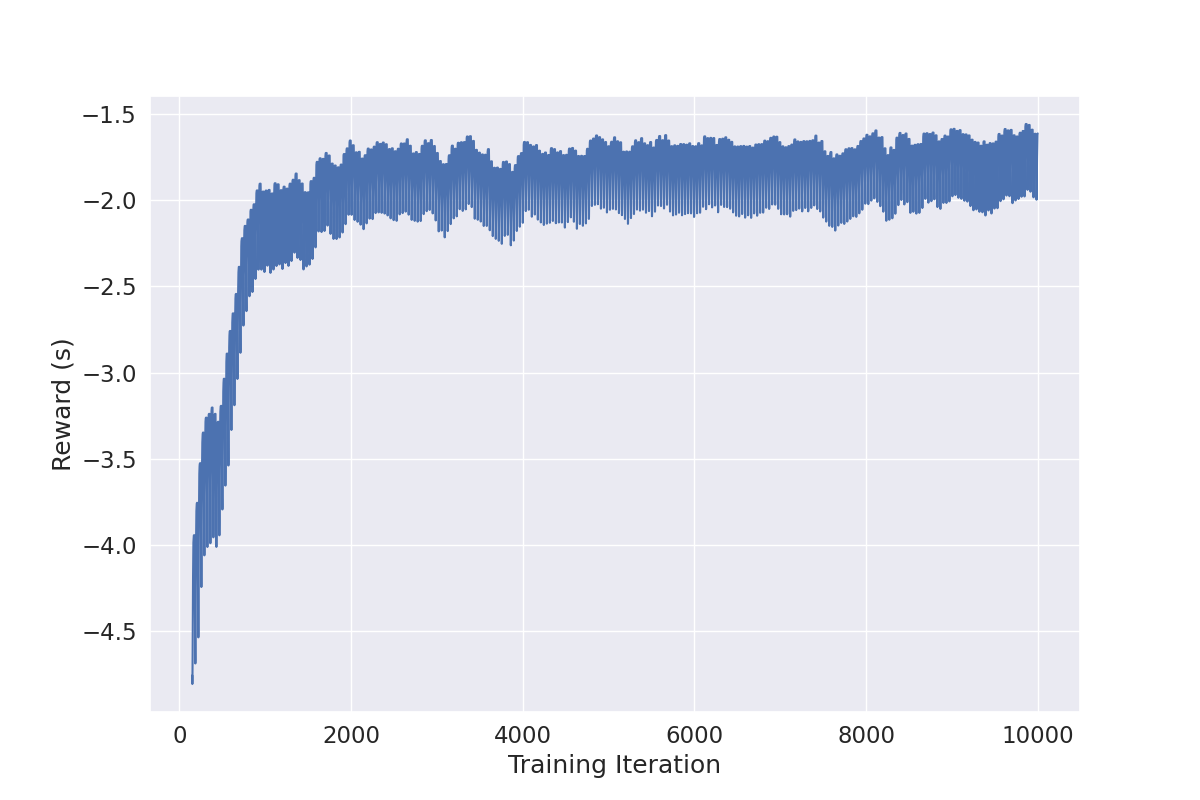}
        \caption{\(\rho_{\text{sep}}=0.95,\ \rho_{\text{noise}}=0,\ \rho_{\text{energy}}=0.05\)}
        \label{fig:training-energy-separation}
    \end{subfigure}

    \caption{Training reward curves for representative policies across different reward weightings. Top row: single-objective training (\(\rho=1.0\) for one component). Bottom row: multi-objective policies balancing two objectives. All policies are trained over 10,000 iterations.}
    \label{fig:training-curves}
\end{figure*}

\subsection{Training Curves}
\paragraph{Single-Objective Policies}
The top row of \cref{fig:training-curves} shows the training reward curves for single-objective policies, each optimized for a distinct objective. These single-objective policies also serve as reference behaviors for interpreting the multi-objective results, since each one represents an extreme policy that prioritizes only one objective.

The separation-only policy (\cref{fig:training-separation-only}) uses \(\rho_{\text{sep}} = 1.0\), with all other weights set to zero. Since the separation reward is only triggered during potential conflict events, the learning process is slower, with the model requiring more iterations than others to converge to a well-performing policy. However, given sufficient training time, the policy converges to a high-performing solution that maintains strong separation guarantees. As we later show in testing, this corresponds to a low number of loss-of-separation (LOS) events.

The energy-only policy (\cref{fig:training-energy-only}) is governed solely by \(\rho_{\text{energy}} = 1.0\), where the reward directly penalizes altitude increases. The learning dynamics in this setting are immediate: the agent converges within a few episodes to a policy that performs no altitude adjustments, maintaining its initial altitude throughout the flight. This leads to a final reward of zero, indicating complete minimization of energy costs.

The noise-only policy (\cref{fig:training-noise-only}) uses \(\rho_{\text{noise}} = 1.0\), and benefits from a dense reward signal applied at every timestep. The agent quickly identifies the optimal strategy of flying at the highest allowable altitude to minimize perceived ground-level noise. Training convergence is fast, but the final reward shows significant variance due to how the aircraft are initialized in each scenario. Since aircraft begin each episode at the lowest altitude, they initially incur a large noise penalty before ascending, even under optimal behavior.
\begin{figure*}[h!]
    \centering
    \begin{subfigure}[b]{0.32\textwidth}
        \centering
        \scalebox{.75}{\begin{tikzpicture}

\definecolor{darkslategray38}{RGB}{38,38,38}
\definecolor{lavender234234242}{RGB}{234,234,242}
\definecolor{steelblue76114176}{RGB}{76,114,176}

\begin{axis}[
axis background/.style={fill=lavender234234242},
width=6cm, height=6cm,
axis line style={white},
tick align=outside,
x grid style={white},
xlabel=\textcolor{darkslategray38}{Average Noise Increase},
xmajorgrids,
xmajorticks=true,
xmin=20.5736539429203, xmax=24.1,
xtick style={color=darkslategray38},
y grid style={white},
ylabel=\textcolor{darkslategray38}{Average LOS Events},
ymajorgrids,
ymajorticks=true,
ymin=-1.41708654632227, ymax=29.400813645063,
ytick style={color=darkslategray38}
]
\path [draw=blue, thick]
(axis cs:23.6046991257299,-0.00544491975611994)
--(axis cs:23.6046991257299,3.08544491975612);

\path [draw=blue, thick]
(axis cs:21.1174841032787,1.44818848942589)
--(axis cs:21.1174841032787,6.03181151057411);

\path [draw=blue, thick]
(axis cs:20.8384336453849,6.92754170085256)
--(axis cs:20.8384336453849,13.0724582991474);

\path [draw=blue, thick]
(axis cs:20.7120053701213,28)
--(axis cs:20.7120053701213,28);

\path [draw=blue, thick]
(axis cs:20.7433154669447,26.6658337398375)
--(axis cs:20.7433154669447,28.3741662601625);

\path [draw=blue, thick]
(axis cs:20.7169324441258,27.1542204861135)
--(axis cs:20.7169324441258,28.4057795138865);

\path [draw=blue, thick]
(axis cs:20.7066746536516,28)
--(axis cs:20.7066746536516,28);

\path [draw=blue, thick]
(axis cs:20.7056148041611,28)
--(axis cs:20.7056148041611,28);

\path [draw=blue, thick]
(axis cs:20.70519913624,28)
--(axis cs:20.70519913624,28);

\path [draw=blue, thick]
(axis cs:20.7048938152846,28)
--(axis cs:20.7048938152846,28);

\path [draw=blue, thick]
(axis cs:20.7048602417347,28)
--(axis cs:20.7048602417347,28);

\addplot [semithick, blue, mark=-, mark size=5, mark options={solid}, only marks]
table {%
23.6046991257299 -0.00544491975611994
21.1174841032787 1.44818848942589
20.8384336453849 6.92754170085256
20.7120053701213 28
20.7433154669447 26.6658337398375
20.7169324441258 27.1542204861135
20.7066746536516 28
20.7056148041611 28
20.70519913624 28
20.7048938152846 28
20.7048602417347 28
};
\addplot [semithick, blue, mark=-, mark size=5, mark options={solid}, only marks]
table {%
23.6046991257299 3.08544491975612
21.1174841032787 6.03181151057411
20.8384336453849 13.0724582991474
20.7120053701213 28
20.7433154669447 28.3741662601625
20.7169324441258 28.4057795138865
20.7066746536516 28
20.7056148041611 28
20.70519913624 28
20.7048938152846 28
20.7048602417347 28
};
\addplot [semithick, steelblue76114176, mark=*, mark size=3, mark options={solid}]
table {%
23.6046991257299 1.54
21.1174841032787 3.74
20.8384336453849 10
20.7120053701213 28
20.7433154669447 27.52
20.7169324441258 27.78
20.7066746536516 28
20.7056148041611 28
20.70519913624 28
20.7048938152846 28
20.7048602417347 28
};
\end{axis}

\end{tikzpicture}}
        \caption{The tradeoff between traffic congestion and noise increase.}
        \label{fig:tradeoff-safety-noise}
    \end{subfigure}
    \hfill
    \begin{subfigure}[b]{0.32\textwidth}
        \centering
        \scalebox{.75}{\begin{tikzpicture}

\definecolor{darkslategray38}{RGB}{38,38,38}
\definecolor{lavender234234242}{RGB}{234,234,242}
\definecolor{steelblue76114176}{RGB}{76,114,176}

\begin{axis}[
axis background/.style={fill=lavender234234242},
width=6cm, height=6cm,
axis line style={white},
tick align=outside,
x grid style={white},
xlabel=\textcolor{darkslategray38}{Average Altitude Adjustments},
xmajorgrids,
xmajorticks=true,
xmin=-0.199305147058824, xmax=4.19996691176471,
xtick style={color=darkslategray38},
y grid style={white},
ylabel=\textcolor{darkslategray38}{Average Noise Increase},
ymajorgrids,
ymajorticks=true,
ymin=20.4391438616328, ymax=27.2072284897121,
ytick style={color=darkslategray38}
]
\addplot [line width=0.9pt, steelblue76114176, mark=*, mark size=4.5, mark options={solid}]
table {%
0.000661764705882353 26.8995882793448
0.00110294117647059 26.8995123090261
0.0265441176470588 26.8959457867238
2.02941176470588 23.0338842837022
2.07419117647059 23.0374524528488
2.93242647058824 22.1955750501913
4 20.746784072
};
\end{axis}

\end{tikzpicture}}
        \caption{The tradeoff between noise increase and average altitude increases.}
        \label{fig:tradeoff-energy-noise}
    \end{subfigure}
    \hfill
    \begin{subfigure}[b]{0.32\textwidth}
        \centering
        \scalebox{.75}{\begin{tikzpicture}

\definecolor{darkslategray38}{RGB}{38,38,38}
\definecolor{lavender234234242}{RGB}{234,234,242}
\definecolor{steelblue76114176}{RGB}{76,114,176}

\begin{axis}[
axis background/.style={fill=lavender234234242},
width=6cm, height=6cm,
axis line style={white},
tick align=outside,
x grid style={white},
xlabel=\textcolor{darkslategray38}{Average Altitude Adjustments},
xmajorgrids,
xmajorticks=true,
xmin=-0.812595588235294, xmax=19.2208308823529,
xtick style={color=darkslategray38},
y grid style={white},
ylabel=\textcolor{darkslategray38}{Average LOS Events},
ymajorgrids,
ymajorticks=true,
ymin=-1.16772603652664, ymax=19.4513214113008,
ytick style={color=darkslategray38}
]
\path [draw=blue, thick]
(axis cs:18.3102205882353,-0.0162729012593024)
--(axis cs:18.3102205882353,3.7362729012593);

\path [draw=blue, thick]
(axis cs:2.57764705882353,0.0573907033213874)
--(axis cs:2.57764705882353,3.20260929667861);

\path [draw=blue, thick]
(axis cs:2.57102941176471,-0.164944443682345)
--(axis cs:2.57102941176471,3.88494444368235);

\path [draw=blue, thick]
(axis cs:1.81257352941176,0.190927796562548)
--(axis cs:1.81257352941176,4.60907220343745);

\path [draw=blue, thick]
(axis cs:2.26632352941176,-0.230496607079937)
--(axis cs:2.26632352941176,2.79049660707994);

\path [draw=blue, thick]
(axis cs:1.81308823529412,0.190024875775822)
--(axis cs:1.81308823529412,4.20997512422418);

\path [draw=blue, thick]
(axis cs:0.120073529411765,17.1659080181459)
--(axis cs:0.120073529411765,18.5140919818541);

\path [draw=blue, thick]
(axis cs:0.0980147058823529,17.3803847577293)
--(axis cs:0.0980147058823529,18.4196152422707);

\addplot [line width=0.9pt, blue, mark=-, mark size=5, mark options={solid}, only marks]
table {%
18.3102205882353 -0.0162729012593024
2.57764705882353 0.0573907033213874
2.57102941176471 -0.164944443682345
1.81257352941176 0.190927796562548
2.26632352941176 -0.230496607079937
1.81308823529412 0.190024875775822
0.120073529411765 17.1659080181459
0.0980147058823529 17.3803847577293
};
\addplot [line width=0.9pt, blue, mark=-, mark size=5, mark options={solid}, only marks]
table {%
18.3102205882353 3.7362729012593
2.57764705882353 3.20260929667861
2.57102941176471 3.88494444368235
1.81257352941176 4.60907220343745
2.26632352941176 2.79049660707994
1.81308823529412 4.20997512422418
0.120073529411765 18.5140919818541
0.0980147058823529 18.4196152422707
};
\addplot [line width=0.9pt, steelblue76114176, mark=*, mark size=4.5, mark options={solid}]
table {%
18.3102205882353 1.86
2.57764705882353 1.63
2.57102941176471 1.86
1.81257352941176 2.4
2.26632352941176 1.28
1.81308823529412 2.2
0.120073529411765 17.84
0.0980147058823529 17.9
};
\end{axis}

\end{tikzpicture}}
        \caption{The tradeoff between LOS events and average altitude increases.}
        \label{fig:tradeoff-energy-separation}
    \end{subfigure}
    \caption{Key tradeoffs across settings: (a) separation vs noise, (b) altitude increases vs noise, and (c) separation vs altitude.}
    \label{fig:combined-key-tradeoffs}
\end{figure*}

\paragraph{Multi-Objective Policies}
The bottom row of \cref{fig:training-curves} shows training curves for policies optimized under multi-objective reward functions. These settings require agents to balance competing priorities, often leading to more complex learning dynamics compared to the single-objective cases.

In the noise–energy setting (\(\rho_{\text{noise}} = 0.1\), \(\rho_{\text{energy}} = 0.9\), \(\rho_{\text{sep}} = 0.0\); \cref{fig:training-noise-energy}), we observe that the model converges to a stable reward around \(-15\). The learning curve initially oscillates around a suboptimal reward for the first 2,000 iterations but then rapidly transitions to a higher-performing policy. This abrupt change is likely due to the directly competing nature of the noise and energy objectives, which forces the agent to explore a range of altitude levels before discovering a stable solution that balances both rewards effectively.

In the noise--separation setting (\(\rho_{\text{noise}} = 0.05\), \(\rho_{\text{energy}} = 0.0\), \(\rho_{\text{sep}} = 0.95\); \cref{fig:training-noise-separation}), we observe a gradual but consistent improvement in reward. Unlike the noise--energy case, where the model quickly converged after a brief period of instability, the learning curve here exhibits a steady upward trend throughout training, eventually plateauing around \(-5\). This behavior reflects the partially aligned nature of the two objectives: increasing altitude simultaneously reduces noise exposure and enhances vertical separation. As a result, the agent can make progress on both objectives concurrently, leading to smoother convergence.

In the separation--energy setting (\(\rho_{\text{sep}} = 0.95\), \(\rho_{\text{energy}} = 0.05\), \(\rho_{\text{noise}} = 0.0\); \cref{fig:training-energy-separation}), we observe a similar learning curve to the noise--separation case. The agent learns a policy that balances safety with low energy expenditure. Since the two objectives are not inherently conflicting, the model is able to improve performance on both simultaneously, resulting in consistent reward improvement.

\begin{figure*}[t!]
    \centering

    \begin{subfigure}[t]{\textwidth}
        \centering
        
        \begin{subfigure}[b]{0.18\textwidth}
            \centering
            \scalebox{.91}{\begin{tikzpicture}

\definecolor{chocolate197870}{RGB}{197,87,0}
\definecolor{darkslategray38}{RGB}{38,38,38}
\definecolor{lavender234234242}{RGB}{234,234,242}

\begin{axis}[
width = 4cm,
height = 4cm,
label style = {font=\footnotesize,},
ylabel style = {yshift=-9pt},
xlabel=\textcolor{darkslategray38}{Altitude Levels ($10^3$ ft)},
xmajorgrids,
xmajorticks=true,
xmin=-0.64, xmax=4.64,
xtick style={color=darkslategray38},
xtick={0,1,2,3,4},
xticklabels={1,1.5,2,2.5,3},
ymajorgrids,
ymajorticks=true,
ymin=0, ymax=1.05,
ytick style={color=darkslategray38}
]
\draw[draw=white,fill=chocolate197870] (axis cs:-0.4,0) rectangle (axis cs:0.4,0.288668153253598);
\draw[draw=white,fill=chocolate197870] (axis cs:0.6,0) rectangle (axis cs:1.4,0.27179201297385);
\draw[draw=white,fill=chocolate197870] (axis cs:1.6,0) rectangle (axis cs:2.4,0.204439489154673);
\draw[draw=white,fill=chocolate197870] (axis cs:2.6,0) rectangle (axis cs:3.4,0.143168457328198);
\draw[draw=white,fill=chocolate197870] (axis cs:3.6,0) rectangle (axis cs:4.4,0.0919318872896817);
\end{axis}

\end{tikzpicture}}
            \caption*{$\rho_{\text{sep}} = 1.0$}
            \label{fig:cn-plot-0}
        \end{subfigure}\hfill
        \begin{subfigure}[b]{0.18\textwidth}
            \centering
            \scalebox{.91}{\begin{tikzpicture}

\definecolor{chocolate197870}{RGB}{197,87,0}
\definecolor{darkslategray38}{RGB}{38,38,38}
\definecolor{lavender234234242}{RGB}{234,234,242}

\begin{axis}[
width = 4cm,
height = 4cm,
label style = {font=\footnotesize,},
ylabel style = {yshift=-9pt},
xlabel=\textcolor{darkslategray38}{Altitude Levels ($10^3$ ft)},
xmajorgrids,
xmajorticks=true,
xmin=-0.64, xmax=4.64,
xtick style={color=darkslategray38},
xtick={0,1,2,3,4},
xticklabels={1,1.5,2,2.5,3},
ymajorgrids,
ymajorticks=true,
ymin=0, ymax=1.05,
ytick style={color=darkslategray38}
]
\draw[draw=white,fill=chocolate197870] (axis cs:-0.4,0) rectangle (axis cs:0.4,0.0783367556468172);
\draw[draw=white,fill=chocolate197870] (axis cs:0.6,0) rectangle (axis cs:1.4,0.0593429158110883);
\draw[draw=white,fill=chocolate197870] (axis cs:1.6,0) rectangle (axis cs:2.4,0.0892197125256673);
\draw[draw=white,fill=chocolate197870] (axis cs:2.6,0) rectangle (axis cs:3.4,0.100667351129363);
\draw[draw=white,fill=chocolate197870] (axis cs:3.6,0) rectangle (axis cs:4.4,0.672433264887064);
\end{axis}

\end{tikzpicture}}
            \caption*{$\rho_{\text{sep}} = 0.95$}
            \label{fig:cn-plot-02}
        \end{subfigure}\hfill
        \begin{subfigure}[b]{0.18\textwidth}
            \centering
            \scalebox{.91}{\begin{tikzpicture}

\definecolor{chocolate197870}{RGB}{197,87,0}
\definecolor{darkslategray38}{RGB}{38,38,38}
\definecolor{lavender234234242}{RGB}{234,234,242}

\begin{axis}[
width = 4cm,
height = 4cm,
label style = {font=\footnotesize,},
ylabel style = {yshift=-9pt},
xlabel=\textcolor{darkslategray38}{Altitude Levels ($10^3$ ft)},
xmajorgrids,
xmajorticks=true,
xmin=-0.64, xmax=4.64,
xtick style={color=darkslategray38},
xtick={0,1,2,3,4},
xticklabels={1,1.5,2,2.5,3},
ymajorgrids,
ymajorticks=true,
ymin=0, ymax=1.05,
ytick style={color=darkslategray38}
]
\draw[draw=white,fill=chocolate197870] (axis cs:-0.4,0) rectangle (axis cs:0.4,0.0733850660362814);
\draw[draw=white,fill=chocolate197870] (axis cs:0.6,0) rectangle (axis cs:1.4,0.036743923120407);
\draw[draw=white,fill=chocolate197870] (axis cs:1.6,0) rectangle (axis cs:2.4,0.0485122565393905);
\draw[draw=white,fill=chocolate197870] (axis cs:2.6,0) rectangle (axis cs:3.4,0.0988231666581016);
\draw[draw=white,fill=chocolate197870] (axis cs:3.6,0) rectangle (axis cs:4.4,0.742535587645819);
\end{axis}

\end{tikzpicture}}
            \caption*{$\rho_{\text{sep}} = 0.9$}
            \label{fig:cn-plot-04}
        \end{subfigure}\hfill
        \begin{subfigure}[b]{0.18\textwidth}
            \centering
            \scalebox{.91}{\begin{tikzpicture}

\definecolor{chocolate197870}{RGB}{197,87,0}
\definecolor{darkslategray38}{RGB}{38,38,38}
\definecolor{lavender234234242}{RGB}{234,234,242}

\begin{axis}[
width = 4cm,
height = 4cm,
label style = {font=\footnotesize,},
ylabel style = {yshift=-9pt},
xlabel=\textcolor{darkslategray38}{Altitude Levels ($10^3$ ft)},
xmajorgrids,
xmajorticks=true,
xmin=-0.64, xmax=4.64,
xtick style={color=darkslategray38},
xtick={0,1,2,3,4},
xticklabels={1,1.5,2,2.5,3},
ymajorgrids,
ymajorticks=true,
ymin=0, ymax=1.05,
ytick style={color=darkslategray38}
]
\draw[draw=white,fill=chocolate197870] (axis cs:-0.4,0) rectangle (axis cs:0.4,0.0724682404978656);
\draw[draw=white,fill=chocolate197870] (axis cs:0.6,0) rectangle (axis cs:1.4,0.0349740266419791);
\draw[draw=white,fill=chocolate197870] (axis cs:1.6,0) rectangle (axis cs:2.4,0.0354369181710641);
\draw[draw=white,fill=chocolate197870] (axis cs:2.6,0) rectangle (axis cs:3.4,0.0376999434243687);
\draw[draw=white,fill=chocolate197870] (axis cs:3.6,0) rectangle (axis cs:4.4,0.819420871264723);
\end{axis}

\end{tikzpicture}}
            \caption*{$\rho_{\text{sep}} = 0.85$}
            \label{fig:cn-plot-06}
        \end{subfigure}\hfill
        \begin{subfigure}[b]{0.18\textwidth}
            \centering
            \scalebox{.91}{\begin{tikzpicture}

\definecolor{chocolate197870}{RGB}{197,87,0}
\definecolor{darkslategray38}{RGB}{38,38,38}
\definecolor{lavender234234242}{RGB}{234,234,242}

\begin{axis}[
width = 4cm,
height = 4cm,
label style = {font=\footnotesize,},
ylabel style = {yshift=-9pt},
xlabel=\textcolor{darkslategray38}{Altitude Levels ($10^3$ ft)},
xmajorgrids,
xmajorticks=true,
xmin=-0.64, xmax=4.64,
xtick style={color=darkslategray38},
xtick={0,1,2,3,4},
xticklabels={1,1.5,2,2.5,3},
ymajorgrids,
ymajorticks=true,
ymin=0, ymax=1.05,
ytick style={color=darkslategray38}
]
\draw[draw=white,fill=chocolate197870] (axis cs:-0.4,0) rectangle (axis cs:0.4,0.0724645134745937);
\draw[draw=white,fill=chocolate197870] (axis cs:0.6,0) rectangle (axis cs:1.4,0.0357436741411232);
\draw[draw=white,fill=chocolate197870] (axis cs:1.6,0) rectangle (axis cs:2.4,0.0402180621271343);
\draw[draw=white,fill=chocolate197870] (axis cs:2.6,0) rectangle (axis cs:3.4,0.0428409792223822);
\draw[draw=white,fill=chocolate197870] (axis cs:3.6,0) rectangle (axis cs:4.4,0.808732771034766);
\end{axis}

\end{tikzpicture}}
            \caption*{$\rho_{\text{sep}} = 0.0$}
            \label{fig:cn-plot-09}
        \end{subfigure}

        \caption{Altitude distribution plots for separation--noise tradeoffs. As noise weighting increases (i.e., $\rho_{\text{sep}}$ decreases), aircraft increasingly cluster at higher altitudes.}
        \label{fig:row-noise-sep}
    \end{subfigure}

    \vspace{0.8em}

    \begin{subfigure}[t]{\textwidth}
        \centering

        \begin{subfigure}[b]{0.18\textwidth}
            \centering
            \scalebox{.91}{\begin{tikzpicture}

\definecolor{chocolate197870}{RGB}{197,87,0}
\definecolor{darkslategray38}{RGB}{38,38,38}
\definecolor{lavender234234242}{RGB}{234,234,242}

\begin{axis}[
width = 4cm,
height = 4cm,
label style = {font=\footnotesize,},
ylabel style = {yshift=-9pt},
xlabel=\textcolor{darkslategray38}{Altitude Levels ($10^3$ ft)},
xmajorgrids,
xmajorticks=true,
xmin=-0.64, xmax=4.64,
xtick style={color=darkslategray38},
xtick={0,1,2,3,4},
xticklabels={1,1.5,2,2.5,3},
ymajorgrids,
ymajorticks=true,
ymin=0, ymax=1.05,
ytick style={color=darkslategray38}
]
\draw[draw=white,fill=chocolate197870] (axis cs:-0.4,0) rectangle (axis cs:0.4,1);
\end{axis}

\end{tikzpicture}}
            \caption*{$\rho_{\text{energy}} = 1.0$}
            \label{fig:cn-noise-energy-1}
        \end{subfigure}\hfill
        \begin{subfigure}[b]{0.18\textwidth}
            \centering
            \scalebox{.91}{\begin{tikzpicture}

\definecolor{chocolate197870}{RGB}{197,87,0}
\definecolor{darkslategray38}{RGB}{38,38,38}
\definecolor{lavender234234242}{RGB}{234,234,242}

\begin{axis}[
width = 4cm,
height = 4cm,
label style = {font=\footnotesize,},
ylabel style = {yshift=-9pt},
xlabel=\textcolor{darkslategray38}{Altitude Levels ($10^3$ ft)},
xmajorgrids,
xmajorticks=true,
xmin=-0.64, xmax=4.64,
xtick style={color=darkslategray38},
xtick={0,1,2,3,4},
xticklabels={1,1.5,2,2.5,3},
ymajorgrids,
ymajorticks=true,
ymin=0, ymax=1.05,
ytick style={color=darkslategray38}
]
\draw[draw=white,fill=chocolate197870] (axis cs:-0.4,0) rectangle (axis cs:0.4,0.0742453501372091);
\draw[draw=white,fill=chocolate197870] (axis cs:0.6,0) rectangle (axis cs:1.4,0.0715011688179693);
\draw[draw=white,fill=chocolate197870] (axis cs:1.6,0) rectangle (axis cs:2.4,0.807704034962903);
\draw[draw=white,fill=chocolate197870] (axis cs:2.6,0) rectangle (axis cs:3.4,0.0465494460819189);
\end{axis}

\end{tikzpicture}}
            \caption*{$\rho_{\text{energy}} = 0.9$}
            \label{fig:cn-noise-energy-09}
        \end{subfigure}\hfill
        \begin{subfigure}[b]{0.18\textwidth}
            \centering
            \scalebox{.91}{\begin{tikzpicture}

\definecolor{chocolate197870}{RGB}{197,87,0}
\definecolor{darkslategray38}{RGB}{38,38,38}
\definecolor{lavender234234242}{RGB}{234,234,242}

\begin{axis}[
width = 4cm,
height = 4cm,
label style = {font=\footnotesize,},
ylabel style = {yshift=-9pt},
xlabel=\textcolor{darkslategray38}{Altitude Levels ($10^3$ ft)},
xmajorgrids,
xmajorticks=true,
xmin=-0.64, xmax=4.64,
xtick style={color=darkslategray38},
xtick={0,1,2,3,4},
xticklabels={1,1.5,2,2.5,3},
ymajorgrids,
ymajorticks=true,
ymin=0, ymax=1.05,
ytick style={color=darkslategray38}
]
\draw[draw=white,fill=chocolate197870] (axis cs:-0.4,0) rectangle (axis cs:0.4,0.0725954666122116);
\draw[draw=white,fill=chocolate197870] (axis cs:0.6,0) rectangle (axis cs:1.4,0.0457933428629773);
\draw[draw=white,fill=chocolate197870] (axis cs:1.6,0) rectangle (axis cs:2.4,0.291607106391668);
\draw[draw=white,fill=chocolate197870] (axis cs:2.6,0) rectangle (axis cs:3.4,0.561721462119665);
\draw[draw=white,fill=chocolate197870] (axis cs:3.6,0) rectangle (axis cs:4.4,0.0282826220134776);
\end{axis}

\end{tikzpicture}}
            \caption*{$\rho_{\text{energy}} = 0.875$}
            \label{fig:cn-noise-energy-0875}
        \end{subfigure}\hfill
        \begin{subfigure}[b]{0.18\textwidth}
            \centering
            \scalebox{.91}{\begin{tikzpicture}

\definecolor{chocolate197870}{RGB}{197,87,0}
\definecolor{darkslategray38}{RGB}{38,38,38}
\definecolor{lavender234234242}{RGB}{234,234,242}

\begin{axis}[
width = 4cm,
height = 4cm,
label style = {font=\footnotesize,},
ylabel style = {yshift=-9pt},
xlabel=\textcolor{darkslategray38}{Altitude Levels ($10^3$ ft)},
xmajorgrids,
xmajorticks=true,
xmin=-0.64, xmax=4.64,
xtick style={color=darkslategray38},
xtick={0,1,2,3,4},
xticklabels={1,1.5,2,2.5,3},
ymajorgrids,
ymajorticks=true,
ymin=0, ymax=1.05,
ytick style={color=darkslategray38}
]
\draw[draw=white,fill=chocolate197870] (axis cs:-0.4,0) rectangle (axis cs:0.4,0.0728684562377867);
\draw[draw=white,fill=chocolate197870] (axis cs:0.6,0) rectangle (axis cs:1.4,0.0366656381775172);
\draw[draw=white,fill=chocolate197870] (axis cs:1.6,0) rectangle (axis cs:2.4,0.0402139257430834);
\draw[draw=white,fill=chocolate197870] (axis cs:2.6,0) rectangle (axis cs:3.4,0.0418080839247146);
\draw[draw=white,fill=chocolate197870] (axis cs:3.6,0) rectangle (axis cs:4.4,0.808443895916898);
\end{axis}

\end{tikzpicture}}
            \caption*{$\rho_{\text{energy}} = 0.85$}
            \label{fig:cn-noise-energy-085}
        \end{subfigure}\hfill
        \begin{subfigure}[b]{0.18\textwidth}
            \centering
            \scalebox{.91}{\begin{tikzpicture}

\definecolor{chocolate197870}{RGB}{197,87,0}
\definecolor{darkslategray38}{RGB}{38,38,38}
\definecolor{lavender234234242}{RGB}{234,234,242}

\begin{axis}[
width = 4cm,
height = 4cm,
label style = {font=\footnotesize,},
ylabel style = {yshift=-9pt},
xlabel=\textcolor{darkslategray38}{Altitude Levels ($10^3$ ft)},
xmajorgrids,
xmajorticks=true,
xmin=-0.64, xmax=4.64,
xtick style={color=darkslategray38},
xtick={0,1,2,3,4},
xticklabels={1,1.5,2,2.5,3},
ymajorgrids,
ymajorticks=true,
ymin=0, ymax=1.05,
ytick style={color=darkslategray38}
]
\draw[draw=white,fill=chocolate197870] (axis cs:-0.4,0) rectangle (axis cs:0.4,0.0724645134745937);
\draw[draw=white,fill=chocolate197870] (axis cs:0.6,0) rectangle (axis cs:1.4,0.0357436741411232);
\draw[draw=white,fill=chocolate197870] (axis cs:1.6,0) rectangle (axis cs:2.4,0.0402180621271343);
\draw[draw=white,fill=chocolate197870] (axis cs:2.6,0) rectangle (axis cs:3.4,0.0428409792223822);
\draw[draw=white,fill=chocolate197870] (axis cs:3.6,0) rectangle (axis cs:4.4,0.808732771034766);
\end{axis}

\end{tikzpicture}}
            \caption*{$\rho_{\text{energy}} = 0.0$}
            \label{fig:cn-noise-energy-00}
        \end{subfigure}

        \caption{Representative altitude distribution plots for noise--energy tradeoffs. Higher noise prioritization ($\rho_{\text{energy}}$ decreasing) leads to increased clustering at higher altitudes.}
        \label{fig:row-noise-energy}
    \end{subfigure}

    \vspace{0.8em}

    \begin{subfigure}[t]{\textwidth}
        \centering

        \begin{subfigure}[b]{0.18\textwidth}
            \centering
            \scalebox{.91}{\begin{tikzpicture}

\definecolor{chocolate197870}{RGB}{197,87,0}
\definecolor{darkslategray38}{RGB}{38,38,38}
\definecolor{lavender234234242}{RGB}{234,234,242}

\begin{axis}[
width = 4cm,
height = 4cm,
label style = {font=\footnotesize,},
ylabel style = {yshift=-9pt},
xlabel=\textcolor{darkslategray38}{Altitude Levels ($10^3$ ft)},
xmajorgrids,
xmajorticks=true,
xmin=-0.64, xmax=4.64,
xtick style={color=darkslategray38},
xtick={0,1,2,3,4},
xticklabels={1,1.5,2,2.5,3},
ymajorgrids,
ymajorticks=true,
ymin=0, ymax=1.05,
ytick style={color=darkslategray38}
]
\draw[draw=white,fill=chocolate197870] (axis cs:-0.4,0) rectangle (axis cs:0.4,0.253923909178646);
\draw[draw=white,fill=chocolate197870] (axis cs:0.6,0) rectangle (axis cs:1.4,0.24498399959364);
\draw[draw=white,fill=chocolate197870] (axis cs:1.6,0) rectangle (axis cs:2.4,0.189465129273124);
\draw[draw=white,fill=chocolate197870] (axis cs:2.6,0) rectangle (axis cs:3.4,0.184690404835678);
\draw[draw=white,fill=chocolate197870] (axis cs:3.6,0) rectangle (axis cs:4.4,0.126936557118911);
\end{axis}

\end{tikzpicture}}
            \caption*{$\rho_{\text{sep}} = 1.0$}
            \label{fig:ce-plot-00}
        \end{subfigure}\hfill
        \begin{subfigure}[b]{0.18\textwidth}
            \centering
            \scalebox{.91}{\begin{tikzpicture}

\definecolor{chocolate197870}{RGB}{197,87,0}
\definecolor{darkslategray38}{RGB}{38,38,38}
\definecolor{lavender234234242}{RGB}{234,234,242}

\begin{axis}[
width = 4cm,
height = 4cm,
label style = {font=\footnotesize,},
ylabel style = {yshift=-9pt},
xlabel=\textcolor{darkslategray38}{Altitude Levels ($10^3$ ft)},
xmajorgrids,
xmajorticks=true,
xmin=-0.64, xmax=4.64,
xtick style={color=darkslategray38},
xtick={0,1,2,3,4},
xticklabels={1,1.5,2,2.5,3},
ymajorgrids,
ymajorticks=true,
ymin=0, ymax=1.05,
ytick style={color=darkslategray38}
]
\draw[draw=white,fill=chocolate197870] (axis cs:-0.4,0) rectangle (axis cs:0.4,0.785520088611419);
\draw[draw=white,fill=chocolate197870] (axis cs:0.6,0) rectangle (axis cs:1.4,0.0734065048836975);
\draw[draw=white,fill=chocolate197870] (axis cs:1.6,0) rectangle (axis cs:2.4,0.049642533481019);
\draw[draw=white,fill=chocolate197870] (axis cs:2.6,0) rectangle (axis cs:3.4,0.0426442452925184);
\draw[draw=white,fill=chocolate197870] (axis cs:3.6,0) rectangle (axis cs:4.4,0.0487866277313463);
\end{axis}

\end{tikzpicture}}
            \caption*{$\rho_{\text{sep}} = 0.9$}
            \label{fig:ce-plot-02}
        \end{subfigure}\hfill
        \begin{subfigure}[b]{0.18\textwidth}
            \centering
            \scalebox{.91}{\begin{tikzpicture}

\definecolor{chocolate197870}{RGB}{197,87,0}
\definecolor{darkslategray38}{RGB}{38,38,38}
\definecolor{lavender234234242}{RGB}{234,234,242}

\begin{axis}[
width = 4cm,
height = 4cm,
label style = {font=\footnotesize,},
ylabel style = {yshift=-9pt},
xlabel=\textcolor{darkslategray38}{Altitude Levels ($10^3$ ft)},
xmajorgrids,
xmajorticks=true,
xmin=-0.64, xmax=4.64,
xtick style={color=darkslategray38},
xtick={0,1,2,3,4},
xticklabels={1,1.5,2,2.5,3},
ymajorgrids,
ymajorticks=true,
ymin=0, ymax=1.05,
ytick style={color=darkslategray38}
]
\draw[draw=white,fill=chocolate197870] (axis cs:-0.4,0) rectangle (axis cs:0.4,0.80420671262517);
\draw[draw=white,fill=chocolate197870] (axis cs:0.6,0) rectangle (axis cs:1.4,0.0710511749610024);
\draw[draw=white,fill=chocolate197870] (axis cs:1.6,0) rectangle (axis cs:2.4,0.0559049967292306);
\draw[draw=white,fill=chocolate197870] (axis cs:2.6,0) rectangle (axis cs:3.4,0.0413626528455694);
\draw[draw=white,fill=chocolate197870] (axis cs:3.6,0) rectangle (axis cs:4.4,0.0274744628390278);
\end{axis}

\end{tikzpicture}}
            \caption*{$\rho_{\text{sep}} = 0.8$}
            \label{fig:ce-plot-04}
        \end{subfigure}\hfill
        \begin{subfigure}[b]{0.18\textwidth}
            \centering
            \scalebox{.91}{\begin{tikzpicture}

\definecolor{chocolate197870}{RGB}{197,87,0}
\definecolor{darkslategray38}{RGB}{38,38,38}
\definecolor{lavender234234242}{RGB}{234,234,242}

\begin{axis}[
width = 4cm,
height = 4cm,
label style = {font=\footnotesize,},
ylabel style = {yshift=-9pt},
xlabel=\textcolor{darkslategray38}{Altitude Levels ($10^3$ ft)},
xmajorgrids,
xmajorticks=true,
xmin=-0.64, xmax=4.64,
xtick style={color=darkslategray38},
xtick={0,1,2,3,4},
xticklabels={1,1.5,2,2.5,3},
ymajorgrids,
ymajorticks=true,
ymin=0, ymax=1.05,
ytick style={color=darkslategray38}
]
\draw[draw=white,fill=chocolate197870] (axis cs:-0.4,0) rectangle (axis cs:0.4,0.994430506773708);
\draw[draw=white,fill=chocolate197870] (axis cs:0.6,0) rectangle (axis cs:1.4,0.00556949322629202);
\end{axis}

\end{tikzpicture}}
            \caption*{$\rho_{\text{sep}} = 0.7$}
            \label{fig:ce-plot-06}
        \end{subfigure}\hfill
        \begin{subfigure}[b]{0.18\textwidth}
            \centering
            \scalebox{.91}{\begin{tikzpicture}

\definecolor{chocolate197870}{RGB}{197,87,0}
\definecolor{darkslategray38}{RGB}{38,38,38}
\definecolor{lavender234234242}{RGB}{234,234,242}

\begin{axis}[
width = 4cm,
height = 4cm,
label style = {font=\footnotesize,},
ylabel style = {yshift=-9pt},
xlabel=\textcolor{darkslategray38}{Altitude Levels ($10^3$ ft)},
xmajorgrids,
xmajorticks=true,
xmin=-0.64, xmax=4.64,
xtick style={color=darkslategray38},
xtick={0,1,2,3,4},
xticklabels={1,1.5,2,2.5,3},
ymajorgrids,
ymajorticks=true,
ymin=0, ymax=1.05,
ytick style={color=darkslategray38}
]
\draw[draw=white,fill=chocolate197870] (axis cs:-0.4,0) rectangle (axis cs:0.4,0.99894641782059);
\draw[draw=white,fill=chocolate197870] (axis cs:0.6,0) rectangle (axis cs:1.4,0.00105358217940999);
\end{axis}

\end{tikzpicture}}
            \caption*{$\rho_{\text{sep}} = 0.55$}
            \label{fig:ce-plot-09}
        \end{subfigure}

        \caption{Representative altitude distribution plots for energy--separation tradeoffs. Increased emphasis on energy minimization (i.e., lower $\rho_{\text{sep}}$) leads to more aircraft remaining at lower altitudes.}
        \label{fig:row-energy-sep}
    \end{subfigure}

    \caption{Altitude distribution across three trade-offs (top to bottom): separation--noise, noise--energy, and energy--separation. Each row shows five representative settings spanning the trade-off.}
    \label{fig:altitude_distributions_three_rows}
\end{figure*}

\subsection{Separation vs. Noise Tradeoff}

\Cref{fig:tradeoff-safety-noise} illustrates the relationship between vertical separation and noise mitigation across policies trained with varying values of \(\rho_{\text{sep}} \in \{1.0, 0.95, \ldots, 0.55\}\), where \(\rho_{\text{noise}} = 1 - \rho_{\text{sep}}\) and \(\rho_{\text{energy}} = 0\). Over a broad range of weightings, the two objectives are not strictly antagonistic: many policies achieve both low cumulative noise impact and a low number of loss-of-separation (LOS) events. \Cref{fig:row-noise-sep} and \Cref{fig:noise_impact_box_sep_noise} provide further insight by showing aircraft altitude distributions and cumulative noise impact, respectively. When separation is prioritized, aircraft distribute themselves evenly across altitude levels to minimize congestion, but as noise becomes more heavily weighted, they increasingly cluster at the highest altitude (3{,}000 ft) to reduce perceived ground noise. For moderate noise weightings, this behavior remains adaptive. Aircraft ascend to minimize noise, then selectively descend to avoid conflicts before returning to higher altitudes. Once \(\rho_{\text{noise}}\) exceeds a certain threshold (\(\rho_{\text{sep}} \leq 0.8\)), noise begins to dominate decision-making, leading to policies that remain at the highest altitude even in the presence of persistent conflicts, thereby increasing the number of LOS events.

\subsection{Noise vs. Energy Tradeoff}

\Cref{fig:tradeoff-energy-noise} illustrates the relationship between cumulative noise impact and total energy consumption across policies trained under varying objective weightings. Each point corresponds to a different value of \(\rho_{\text{energy}}\), with \(\rho_{\text{noise}} = 1 - \rho_{\text{energy}}\) and \(\rho_{\text{sep}} = 0\). In contrast to the noise–separation case, we observe a much stricter tradeoff: improvements in noise reduction consistently lead to increased energy consumption, and vice versa. \Cref{fig:row-noise-energy} and \Cref{fig:noise_impact_box_sep_energy} show how this tradeoff manifests in both aircraft behavior and overall system performance. When energy is prioritized, aircraft remain at the lowest altitude level (1{,}000 ft) throughout the episode, minimizing the cost of climbing. As noise becomes more heavily weighted, aircraft increasingly occupy higher altitudes (2{,}000 ft and 3{,}000 ft), which reduce ground-level exposure but require additional energy. Policies in this regime tend to commit to a single altitude, and tradeoffs between noise and energy emerge as a hard constraint.

\begin{figure*}[hbt!]
    \centering
    \begin{subfigure}[b]{0.48\textwidth}
        \centering
        \scalebox{0.55}{\begin{tikzpicture}

\definecolor{darkcyan19136136}{RGB}{19,136,136}
\definecolor{darkolivegreen9810715}{RGB}{98,107,15}
\definecolor{darkseagreen156193156}{RGB}{156,193,156}
\definecolor{darkslategray36}{RGB}{36,36,36}
\definecolor{darkslategray38}{RGB}{38,38,38}
\definecolor{gray113122129}{RGB}{113,122,129}
\definecolor{gray129122113}{RGB}{129,122,113}
\definecolor{lavender234234242}{RGB}{234,234,242}
\definecolor{rosybrown187150112}{RGB}{187,150,112}
\definecolor{sienna1678923}{RGB}{167,89,23}
\definecolor{slateblue112112187}{RGB}{112,112,187}
\definecolor{yellowgreen14217524}{RGB}{142,175,24}

\begin{axis}[
width=12cm,
axis background/.style={fill=lavender234234242},
tick align=outside,
x grid style={white},
xlabel=\textcolor{darkslategray38}{Noise Tradeoff Parameter Value $\rho_{\text{noise}}$},
xmajorticks=true,
xmin=-0.5, xmax=9.5,
xtick style={color=darkslategray38},
xtick={0,1,2,3,4,5,6,7,8,9}, %
xticklabels={0.0, 0.025, 0.05, 0.75, 0.1, 0.125, 0.15, 0.175, 0.2, 1.0},
y grid style={white},
ylabel=\textcolor{darkslategray38}{Noise Increase (dB)},
ymajorgrids,
ymajorticks=true,
ymin=10.2935489417384, ymax=37.707355209674,
ytick style={color=darkslategray38}
]
\path [draw=darkslategray36, fill=white, thick]
(axis cs:-0.3,20.2824559375727)
--(axis cs:0.3,20.2824559375727)
--(axis cs:0.3,27.4044714589831)
--(axis cs:-0.3,27.4044714589831)
--(axis cs:-0.3,20.2824559375727)
--cycle;
\addplot [thick, darkslategray36]
table {%
0 20.2824559375727
0 13.9455601769722
};
\addplot [thick, darkslategray36]
table {%
0 27.4044714589831
0 33.4417574348797
};
\addplot [thick, darkslategray36]
table {%
-0.15 13.9455601769722
0.15 13.9455601769722
};
\addplot [thick, darkslategray36]
table {%
-0.15 33.4417574348797
0.15 33.4417574348797
};
\path [draw=darkslategray36, fill=yellowgreen14217524, thick]
(axis cs:0.7,17.5768967379118)
--(axis cs:1.3,17.5768967379118)
--(axis cs:1.3,25.1626331010236)
--(axis cs:0.7,25.1626331010236)
--(axis cs:0.7,17.5768967379118)
--cycle;
\addplot [thick, darkslategray36]
table {%
1 17.5768967379118
1 12.4865052640653
};
\addplot [thick, darkslategray36]
table {%
1 25.1626331010236
1 30.1086386105144
};
\addplot [thick, darkslategray36]
table {%
0.85 12.4865052640653
1.15 12.4865052640653
};
\addplot [thick, darkslategray36]
table {%
0.85 30.1086386105144
1.15 30.1086386105144
};
\path [draw=darkslategray36, fill=darkcyan19136136, thick]
(axis cs:1.7,17.5747838025873)
--(axis cs:2.3,17.5747838025873)
--(axis cs:2.3,24.9420949319826)
--(axis cs:1.7,24.9420949319826)
--(axis cs:1.7,17.5747838025873)
--cycle;
\addplot [thick, darkslategray36]
table {%
2 17.5747838025873
2 12.0752293880884
};
\addplot [thick, darkslategray36]
table {%
2 24.9420949319826
2 29.924306987521
};
\addplot [thick, darkslategray36]
table {%
1.85 12.0752293880884
2.15 12.0752293880884
};
\addplot [thick, darkslategray36]
table {%
1.85 29.924306987521
2.15 29.924306987521
};
\path [draw=darkslategray36, fill=gray113122129, thick]
(axis cs:2.7,17.5300879748325)
--(axis cs:3.3,17.5300879748325)
--(axis cs:3.3,24.8638526098098)
--(axis cs:2.7,24.8638526098098)
--(axis cs:2.7,17.5300879748325)
--cycle;
\addplot [thick, darkslategray36]
table {%
3 17.5300879748325
3 11.5570945647754
};
\addplot [thick, darkslategray36]
table {%
3 24.8638526098098
3 29.8680488711338
};
\addplot [thick, darkslategray36]
table {%
2.85 11.5570945647754
3.15 11.5570945647754
};
\addplot [thick, darkslategray36]
table {%
2.85 29.8680488711338
3.15 29.8680488711338
};
\path [draw=darkslategray36, fill=gray129122113, thick]
(axis cs:3.7,17.5211758919159)
--(axis cs:4.3,17.5211758919159)
--(axis cs:4.3,24.8649839514238)
--(axis cs:3.7,24.8649839514238)
--(axis cs:3.7,17.5211758919159)
--cycle;
\addplot [thick, darkslategray36]
table {%
4 17.5211758919159
4 12.039151224989
};
\addplot [thick, darkslategray36]
table {%
4 24.8649839514238
4 29.8693136611821
};
\addplot [thick, darkslategray36]
table {%
3.85 12.039151224989
4.15 12.039151224989
};
\addplot [thick, darkslategray36]
table {%
3.85 29.8693136611821
4.15 29.8693136611821
};
\path [draw=darkslategray36, fill=darkolivegreen9810715, thick]
(axis cs:4.7,17.5274913965844)
--(axis cs:5.3,17.5274913965844)
--(axis cs:5.3,24.8630808512644)
--(axis cs:4.7,24.8630808512644)
--(axis cs:4.7,17.5274913965844)
--cycle;
\addplot [thick, darkslategray36]
table {%
5 17.5274913965844
5 11.6129128399498
};
\addplot [thick, darkslategray36]
table {%
5 24.8630808512644
5 29.8680488711338
};
\addplot [thick, darkslategray36]
table {%
4.85 11.6129128399498
5.15 11.6129128399498
};
\addplot [thick, darkslategray36]
table {%
4.85 29.8680488711338
5.15 29.8680488711338
};
\path [draw=darkslategray36, fill=sienna1678923, thick]
(axis cs:5.7,17.518062668936)
--(axis cs:6.3,17.518062668936)
--(axis cs:6.3,24.8421606394293)
--(axis cs:5.7,24.8421606394293)
--(axis cs:5.7,17.518062668936)
--cycle;
\addplot [thick, darkslategray36]
table {%
6 17.518062668936
6 11.5435733099872
};
\addplot [thick, darkslategray36]
table {%
6 24.8421606394293
6 29.8715426239738
};
\addplot [thick, darkslategray36]
table {%
5.85 11.5435733099872
6.15 11.5435733099872
};
\addplot [thick, darkslategray36]
table {%
5.85 29.8715426239738
6.15 29.8715426239738
};
\path [draw=darkslategray36, fill=rosybrown187150112, thick]
(axis cs:6.7,17.5164003735566)
--(axis cs:7.3,17.5164003735566)
--(axis cs:7.3,24.8366505587896)
--(axis cs:6.7,24.8366505587896)
--(axis cs:6.7,17.5164003735566)
--cycle;
\addplot [thick, darkslategray36]
table {%
7 17.5164003735566
7 11.5396310448264
};
\addplot [thick, darkslategray36]
table {%
7 24.8366505587896
7 29.8680488711338
};
\addplot [thick, darkslategray36]
table {%
6.85 11.5396310448264
7.15 11.5396310448264
};
\addplot [thick, darkslategray36]
table {%
6.85 29.8680488711338
7.15 29.8680488711338
};
\path [draw=darkslategray36, fill=darkseagreen156193156, thick]
(axis cs:7.7,17.5160387091026)
--(axis cs:8.3,17.5160387091026)
--(axis cs:8.3,24.8352436837571)
--(axis cs:7.7,24.8352436837571)
--(axis cs:7.7,17.5160387091026)
--cycle;
\addplot [thick, darkslategray36]
table {%
8 17.5160387091026
8 11.5396310448264
};
\addplot [thick, darkslategray36]
table {%
8 24.8352436837571
8 29.8680488711338
};
\addplot [thick, darkslategray36]
table {%
7.85 11.5396310448264
8.15 11.5396310448264
};
\addplot [thick, darkslategray36]
table {%
7.85 29.8680488711338
8.15 29.8680488711338
};
\path [draw=darkslategray36, fill=slateblue112112187, thick]
(axis cs:8.7,17.5151178597766)
--(axis cs:9.3,17.5151178597766)
--(axis cs:9.3,24.8328876664419)
--(axis cs:8.7,24.8328876664419)
--(axis cs:8.7,17.5151178597766)
--cycle;
\addplot [thick, darkslategray36]
table {%
9 17.5151178597766
9 11.5396310448264
};
\addplot [thick, darkslategray36]
table {%
9 24.8328876664419
9 29.8680488711338
};
\addplot [thick, darkslategray36]
table {%
8.85 11.5396310448264
9.15 11.5396310448264
};
\addplot [thick, darkslategray36]
table {%
8.85 29.8680488711338
9.15 29.8680488711338
};
\addplot [thick, darkslategray36]
table {%
-0.3 23.3259503431658
0.3 23.3259503431658
};
\addplot [thick, darkslategray36]
table {%
0.7 21.2710722321434
1.3 21.2710722321434
};
\addplot [thick, darkslategray36]
table {%
1.7 20.7383635350197
2.3 20.7383635350197
};
\addplot [thick, darkslategray36]
table {%
2.7 20.315433437379
3.3 20.315433437379
};
\addplot [thick, darkslategray36]
table {%
3.7 20.3170042062085
4.3 20.3170042062085
};
\addplot [thick, darkslategray36]
table {%
4.7 20.315433437379
5.3 20.315433437379
};
\addplot [thick, darkslategray36]
table {%
5.7 20.315433437379
6.3 20.315433437379
};
\addplot [thick, darkslategray36]
table {%
6.7 20.315433437379
7.3 20.315433437379
};
\addplot [thick, darkslategray36]
table {%
7.7 20.315433437379
8.3 20.315433437379
};
\addplot [thick, darkslategray36]
table {%
8.7 20.315433437379
9.3 20.315433437379
};
\end{axis}

\end{tikzpicture}}
        \caption{Cumulative noise increase over ambient levels across 100 test episodes for varying values of \(\rho_{\text{noise}}\). Higher \(\rho_{\text{noise}}\) values prioritize noise mitigation, leading to reduced and more consistent noise levels.}
        \label{fig:noise_impact_box_sep_noise}
    \end{subfigure}
    \hfill
    \begin{subfigure}[b]{0.48\textwidth}
        \centering
        \scalebox{0.55}{\begin{tikzpicture}

\definecolor{darkcyan19136136}{RGB}{19,136,136}
\definecolor{darkolivegreen9810715}{RGB}{98,107,15}
\definecolor{darkseagreen156193156}{RGB}{156,193,156}
\definecolor{darkslategray36}{RGB}{36,36,36}
\definecolor{darkslategray38}{RGB}{38,38,38}
\definecolor{gray113122129}{RGB}{113,122,129}
\definecolor{gray129122113}{RGB}{129,122,113}
\definecolor{lavender234234242}{RGB}{234,234,242}
\definecolor{rosybrown187150112}{RGB}{187,150,112}
\definecolor{sienna1678923}{RGB}{167,89,23}
\definecolor{slateblue112112187}{RGB}{112,112,187}
\definecolor{yellowgreen14217524}{RGB}{142,175,24}

\begin{axis}[
width=12cm,
axis background/.style={fill=lavender234234242},
tick align=outside,
x grid style={white},
xlabel=\textcolor{darkslategray38}{Noise Tradeoff Parameter Value $\rho_{\text{noise}}$},
xmajorticks=true,
xmin=-0.5, xmax=9.5,
xtick style={color=darkslategray38},
xtick={0,1,2,3,4,5,6,7,8,9}, %
xticklabels={0.0, 0.025, 0.05, 0.75, 0.1, 0.125, 0.15, 0.175, 0.2, 1.0},
y grid style={white},
ylabel=\textcolor{darkslategray38}{Noise Increase (dB)},
ymajorgrids,
ymajorticks=true,
ymin=10.2935489417384, ymax=37.707355209674,
ytick style={color=darkslategray38}
]
\path [draw=darkslategray36, fill=white, thick]
(axis cs:-0.3,23.4702340215542)
--(axis cs:0.3,23.4702340215542)
--(axis cs:0.3,30.7906358650418)
--(axis cs:-0.3,30.7906358650418)
--(axis cs:-0.3,23.4702340215542)
--cycle;
\addplot [thick, darkslategray36]
table {%
0 23.4702340215542
0 16.9778670278312
};
\addplot [thick, darkslategray36]
table {%
0 30.7906358650418
0 36.461273106586
};
\addplot [thick, darkslategray36]
table {%
-0.15 16.9778670278312
0.15 16.9778670278312
};
\addplot [thick, darkslategray36]
table {%
-0.15 36.461273106586
0.15 36.461273106586
};
\path [draw=darkslategray36, fill=yellowgreen14217524, thick]
(axis cs:0.7,23.4702339958997)
--(axis cs:1.3,23.4702339958997)
--(axis cs:1.3,30.7901594871948)
--(axis cs:0.7,30.7901594871948)
--(axis cs:0.7,23.4702339958997)
--cycle;
\addplot [thick, darkslategray36]
table {%
1 23.4702339958997
1 16.9772527614846
};
\addplot [thick, darkslategray36]
table {%
1 30.7901594871948
1 36.4612730925272
};
\addplot [thick, darkslategray36]
table {%
0.85 16.9772527614846
1.15 16.9772527614846
};
\addplot [thick, darkslategray36]
table {%
0.85 36.4612730925272
1.15 36.4612730925272
};
\path [draw=darkslategray36, fill=darkcyan19136136, thick]
(axis cs:1.7,23.4619046211725)
--(axis cs:2.3,23.4619046211725)
--(axis cs:2.3,30.7884519757045)
--(axis cs:1.7,30.7884519757045)
--(axis cs:1.7,23.4619046211725)
--cycle;
\addplot [thick, darkslategray36]
table {%
2 23.4619046211725
2 16.9452949288053
};
\addplot [thick, darkslategray36]
table {%
2 30.7884519757045
2 36.4612729965534
};
\addplot [thick, darkslategray36]
table {%
1.85 16.9452949288053
2.15 16.9452949288053
};
\addplot [thick, darkslategray36]
table {%
1.85 36.4612729965534
2.15 36.4612729965534
};
\path [draw=darkslategray36, fill=gray113122129, thick]
(axis cs:2.7,19.8035816555323)
--(axis cs:3.3,19.8035816555323)
--(axis cs:3.3,27.158905889427)
--(axis cs:2.7,27.158905889427)
--(axis cs:2.7,19.8035816555323)
--cycle;
\addplot [thick, darkslategray36]
table {%
3 19.8035816555323
3 13.4442067939471
};
\addplot [thick, darkslategray36]
table {%
3 27.158905889427
3 32.4687751447907
};
\addplot [thick, darkslategray36]
table {%
2.85 13.4442067939471
3.15 13.4442067939471
};
\addplot [thick, darkslategray36]
table {%
2.85 32.4687751447907
3.15 32.4687751447907
};
\path [draw=darkslategray36, fill=gray129122113, thick]
(axis cs:3.7,19.8055711653469)
--(axis cs:4.3,19.8055711653469)
--(axis cs:4.3,27.3123698054208)
--(axis cs:3.7,27.3123698054208)
--(axis cs:3.7,19.8055711653469)
--cycle;
\addplot [thick, darkslategray36]
table {%
4 19.8055711653469
4 13.4809557695692
};
\addplot [thick, darkslategray36]
table {%
4 27.3123698054208
4 32.4189380905233
};
\addplot [thick, darkslategray36]
table {%
3.85 13.4809557695692
4.15 13.4809557695692
};
\addplot [thick, darkslategray36]
table {%
3.85 32.4189380905233
4.15 32.4189380905233
};
\path [draw=darkslategray36, fill=darkolivegreen9810715, thick]
(axis cs:4.7,19.0240841446615)
--(axis cs:5.3,19.0240841446615)
--(axis cs:5.3,26.8062054384508)
--(axis cs:4.7,26.8062054384508)
--(axis cs:4.7,19.0240841446615)
--cycle;
\addplot [thick, darkslategray36]
table {%
5 19.0240841446615
5 13.026189755049
};
\addplot [thick, darkslategray36]
table {%
5 26.8062054384508
5 31.2610189097368
};
\addplot [thick, darkslategray36]
table {%
4.85 13.026189755049
5.15 13.026189755049
};
\addplot [thick, darkslategray36]
table {%
4.85 31.2610189097368
5.15 31.2610189097368
};
\path [draw=darkslategray36, fill=sienna1678923, thick]
(axis cs:5.7,17.5400629740341)
--(axis cs:6.3,17.5400629740341)
--(axis cs:6.3,25.0019381199389)
--(axis cs:5.7,25.0019381199389)
--(axis cs:5.7,17.5400629740341)
--cycle;
\addplot [thick, darkslategray36]
table {%
6 17.5400629740341
6 11.6204185139677
};
\addplot [thick, darkslategray36]
table {%
6 25.0019381199389
6 29.8727952662566
};
\addplot [thick, darkslategray36]
table {%
5.85 11.6204185139677
6.15 11.6204185139677
};
\addplot [thick, darkslategray36]
table {%
5.85 29.8727952662566
6.15 29.8727952662566
};
\path [draw=darkslategray36, fill=rosybrown187150112, thick]
(axis cs:6.7,17.5400629740341)
--(axis cs:7.3,17.5400629740341)
--(axis cs:7.3,24.9219978688521)
--(axis cs:6.7,24.9219978688521)
--(axis cs:6.7,17.5400629740341)
--cycle;
\addplot [thick, darkslategray36]
table {%
7 17.5400629740341
7 11.5721019191904
};
\addplot [thick, darkslategray36]
table {%
7 24.9219978688521
7 29.8704220686952
};
\addplot [thick, darkslategray36]
table {%
6.85 11.5721019191904
7.15 11.5721019191904
};
\addplot [thick, darkslategray36]
table {%
6.85 29.8704220686952
7.15 29.8704220686952
};
\path [draw=darkslategray36, fill=darkseagreen156193156, thick]
(axis cs:7.7,17.5287710718846)
--(axis cs:8.3,17.5287710718846)
--(axis cs:8.3,24.877754822309)
--(axis cs:7.7,24.877754822309)
--(axis cs:7.7,17.5287710718846)
--cycle;
\addplot [thick, darkslategray36]
table {%
8 17.5287710718846
8 11.5565251199874
};
\addplot [thick, darkslategray36]
table {%
8 24.877754822309
8 29.8680488711338
};
\addplot [thick, darkslategray36]
table {%
7.85 11.5565251199874
8.15 11.5565251199874
};
\addplot [thick, darkslategray36]
table {%
7.85 29.8680488711338
8.15 29.8680488711338
};
\path [draw=darkslategray36, fill=slateblue112112187, thick]
(axis cs:8.7,17.5400629740341)
--(axis cs:9.3,17.5400629740341)
--(axis cs:9.3,25.0323441205575)
--(axis cs:8.7,25.0323441205575)
--(axis cs:8.7,17.5400629740341)
--cycle;
\addplot [thick, darkslategray36]
table {%
9 17.5400629740341
9 11.6388986420349
};
\addplot [thick, darkslategray36]
table {%
9 25.0323441205575
9 29.8685235106461
};
\addplot [thick, darkslategray36]
table {%
8.85 11.6388986420349
9.15 11.6388986420349
};
\addplot [thick, darkslategray36]
table {%
8.85 29.8685235106461
9.15 29.8685235106461
};
\addplot [thick, darkslategray36]
table {%
-0.3 26.9701198408153
0.3 26.9701198408153
};
\addplot [thick, darkslategray36]
table {%
0.7 26.9701198035548
1.3 26.9701198035548
};
\addplot [thick, darkslategray36]
table {%
1.7 26.9694306770484
2.3 26.9694306770484
};
\addplot [thick, darkslategray36]
table {%
2.7 22.8773286153062
3.3 22.8773286153062
};
\addplot [thick, darkslategray36]
table {%
3.7 22.81985962121
4.3 22.81985962121
};
\addplot [thick, darkslategray36]
table {%
4.7 21.8175049190803
5.3 21.8175049190803
};
\addplot [thick, darkslategray36]
table {%
5.7 20.315433437379
6.3 20.315433437379
};
\addplot [thick, darkslategray36]
table {%
6.7 20.315433437379
7.3 20.315433437379
};
\addplot [thick, darkslategray36]
table {%
7.7 20.315433437379
8.3 20.315433437379
};
\addplot [thick, darkslategray36]
table {%
8.7 20.315433437379
9.3 20.315433437379
};
\end{axis}

\end{tikzpicture}}
        \caption{Cumulative noise increase over ambient levels across 100 test episodes for varying values of \(\rho_{\text{noise}}\). Higher \(\rho_{\text{noise}}\) values prioritize noise mitigation, leading to reduced and more consistent noise levels.}
        \label{fig:noise_impact_box_sep_energy}
    \end{subfigure}
    \caption{Comparison of noise impact under (a) noise--separation and (b) noise--energy tradeoffs.}
    \label{fig:noise_impact_box_combined}
\end{figure*}

\subsection{Separation vs. Energy Tradeoff}

\Cref{fig:tradeoff-energy-separation} illustrates the relationship between separation and energy objectives across policies trained with varying reward weightings. Unlike the noise–energy setting, this tradeoff is relatively mild: several models are able to reduce the number of LOS events while also minimizing altitude changes. \Cref{fig:row-energy-sep} shows that when energy is prioritized, aircraft tend to remain at the lowest altitude level (1{,}000 ft) to conserve energy. However, when separation is emphasized, aircraft are willing to perform occasional climbs to reduce local congestion and avoid conflicts. Because noise is not a concern in this setting, agents exhibit greater flexibility: they remain at low altitudes by default and ascend only when necessary. This allows learned policies to achieve safety improvements with minimal energy expenditure, resulting in more balanced behavior than in the noise–energy tradeoff.

\begin{table}[htbp]
    \centering
    \caption{Average loss of separation events $\pm$ standard deviation for models trained on different numbers of training scenarios. We report LOS events in both training scenarios, and the holdout scenarios not used in training.}
    \begin{tabular}{llll}
    \hline
    \begin{tabular}[c]{@{}l@{}}Number of Scenarios\\ Trained On\end{tabular}    & Training Scenarios    & Holdout Scenarios    \\ \hline
    1 & $0.681 \pm 1.215$ & $0.938 \pm 1.553$ \\ 
    5   & $1.340 \pm 1.601$ & $1.787 \pm 1.855$ \\ 
    10 & $1.146 \pm 1.346$ & $1.021 \pm 1.323$ \\ 
    15   & $1.365 \pm 1.820$ & $1.021 \pm 1.554$ \\ 
    20 & $1.021 \pm 1.353$ & --  \\ \hline
    \end{tabular}
    \label{tbl:stochastic_scenarios}
\end{table}

\begin{figure*}[t]
    \centering
    \begin{subfigure}[b]{0.24\textwidth}
        \centering
        \includegraphics[width=\textwidth]{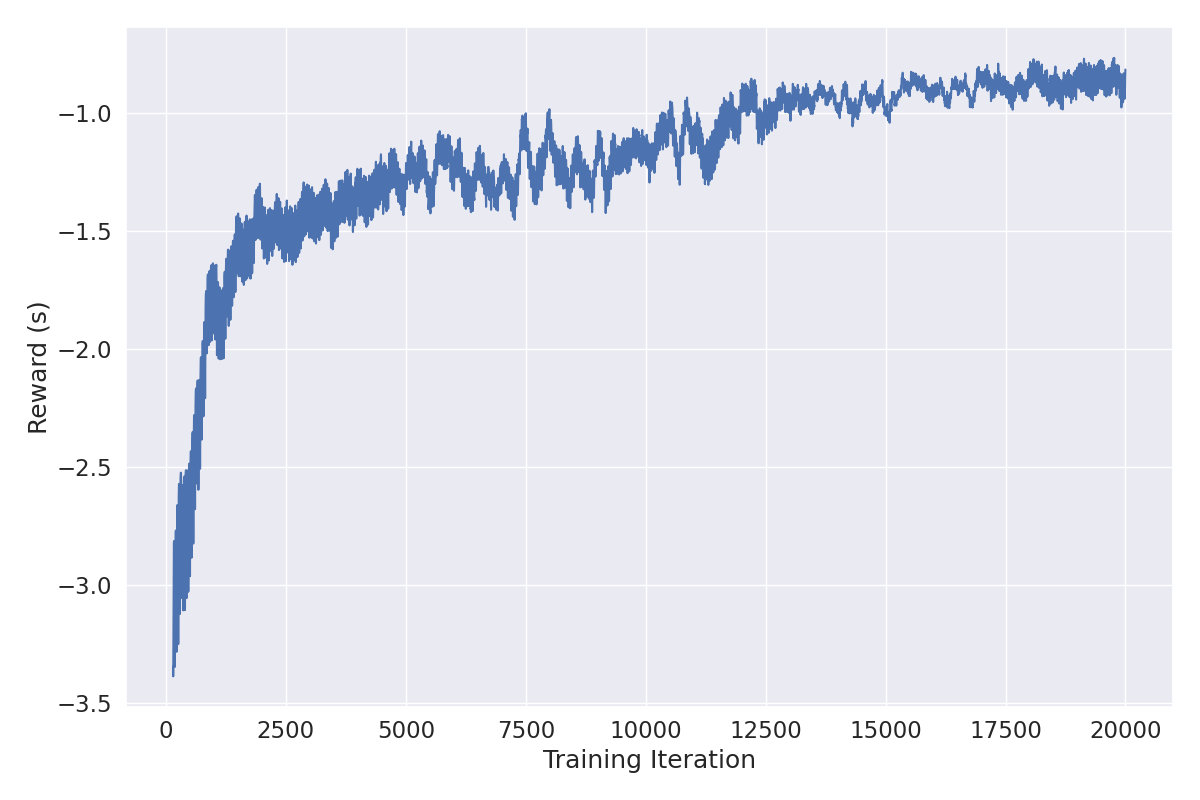}
        \caption{5 Scenarios}
        \label{fig:training-sep-5}
    \end{subfigure}
    \hfill
    \begin{subfigure}[b]{0.24\textwidth}
        \centering
        \includegraphics[width=\textwidth]{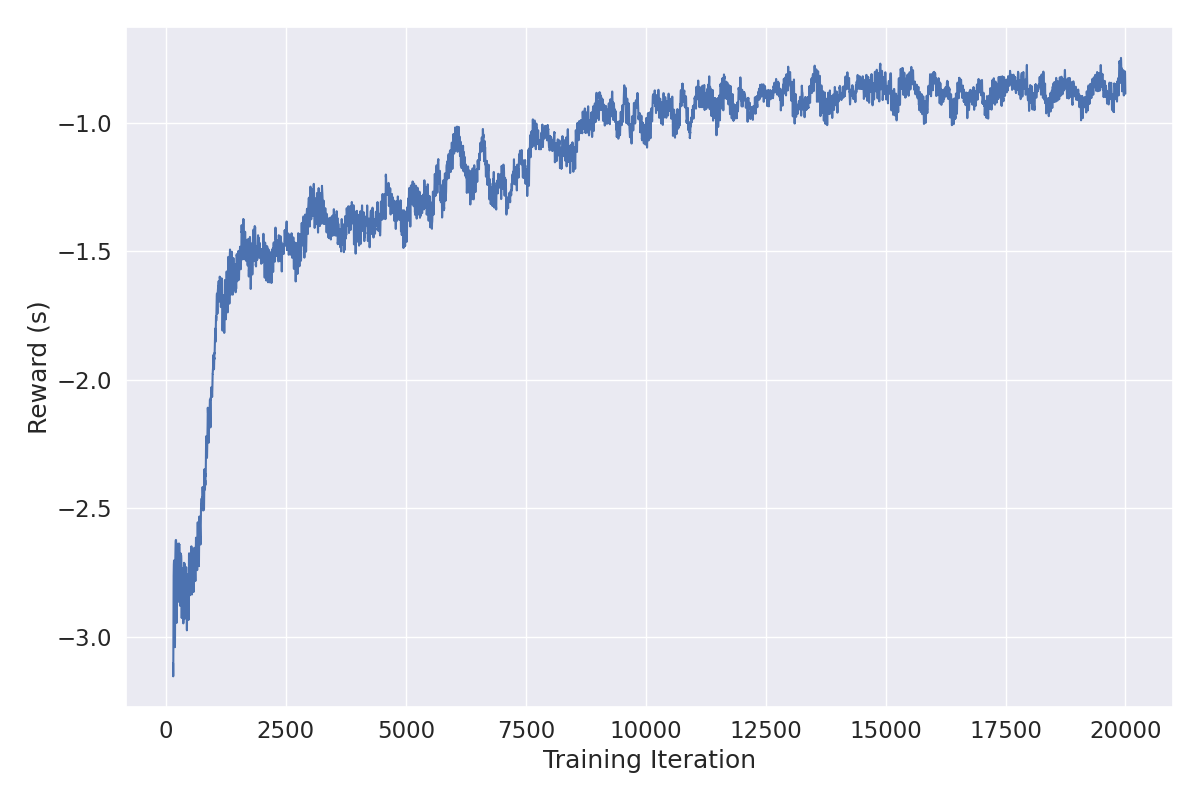}
        \caption{10 Scenarios}
        \label{fig:training-sep-10}
    \end{subfigure}
    \begin{subfigure}[b]{0.24\textwidth}
        \centering
        \includegraphics[width=\textwidth]{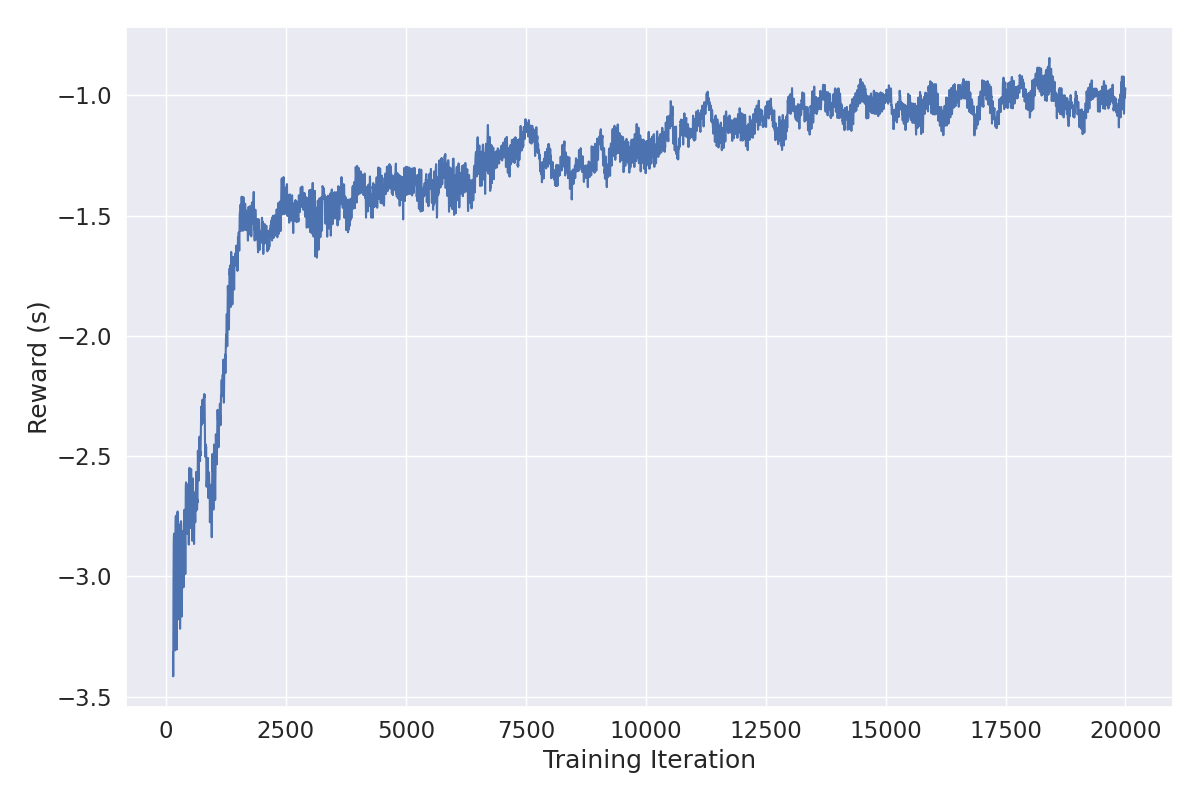}
        \caption{15 Scenarios}
        \label{fig:training-sep-15}
    \end{subfigure}
    \hfill
    \begin{subfigure}[b]{0.24\textwidth}
        \centering
        \includegraphics[width=\textwidth]{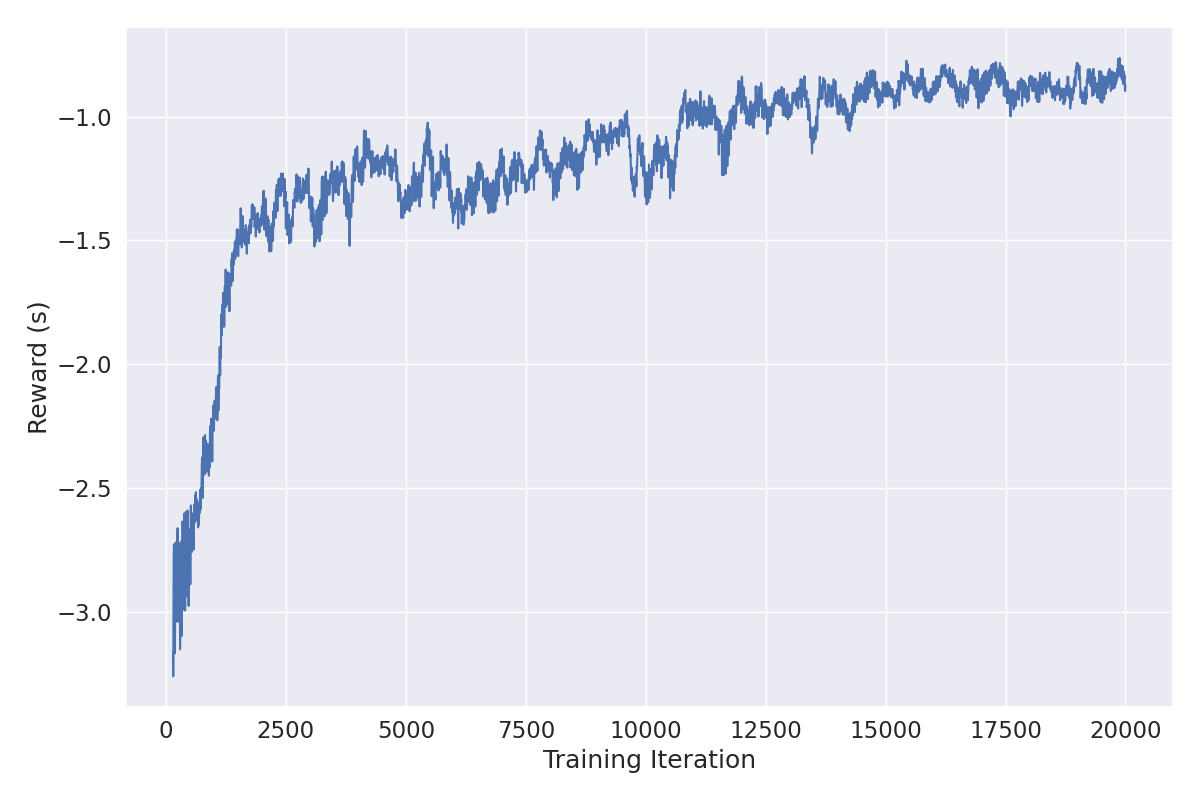}
        \caption{20 Scenarios}
        \label{fig:training-sep-20}
    \end{subfigure}

    \caption{Training reward curves for separation-only policies $\rho_{\text{sep}} = 1$ across different numbers of stochastic traffic scenarios.}
    \label{fig:training-curves-stochastic}
\end{figure*}

\subsection{Generalization Under Stochastic Traffic}

The experiments presented thus far evaluate policies under a deterministic traffic schedule in the South Austin scenario. While this controlled setting enables clear comparisons between reward formulations and policy behaviors, it raises the possibility that learned policies could overfit to a specific traffic realization. To evaluate the robustness of the proposed approach, we conducted additional experiments under stochastic traffic conditions that introduce variability in both traffic demand and departure timing.

To generate stochastic scenarios, we constructed a set of $20$ traffic realizations derived from the nominal demand specification used in the deterministic experiments. For each route, the number of aircraft was sampled from a Gaussian distribution centered on the nominal demand with a standard deviation equal to $25\%$ of the nominal value. The sampled value was rounded to the nearest integer and clipped to ensure feasible aircraft counts, with values restricted to the range $[0,15]$ aircraft per route. Departure schedules were randomized separately for each origin vertiport. For each vertiport, the initial departure time was drawn uniformly from $0$--$60$ s, after which aircraft were released sequentially with inter-departure intervals sampled from a uniform distribution between $120$--$150$ s. This procedure introduces stochasticity in both route demand and departure timing while maintaining safe initial spacing between aircraft.

We trained policies on subsets of these generated scenarios and evaluated performance on held-out scenarios not seen during training. We focus on the separation-only objective for this study, as policies optimizing noise or energy exhibit largely deterministic behavior under the present formulation, whereas separation optimization requires agents to consider interactions with nearby aircraft. Training curves are shown in \cref{fig:training-curves-stochastic}, and policy performance is summarized in \cref{tbl:stochastic_scenarios}.

The results indicate that the reinforcement learning model is able to successfully learn effective policies under stochastic traffic conditions. We see that in \cref{fig:training-sep-5}--\cref{fig:training-sep-20}, the reinforcement learning agent is able to converge to a well-performing policy in all cases. However, training on multiple stochastic scenarios introduces greater variability in the observed rewards, resulting in noisier learning curves and slightly slower convergence compared to the deterministic setting. When training on a small number of scenarios (e.g., $1$ or $5$), we observe a modest reduction in performance on held-out scenarios, indicating limited overfitting to the training realizations. As the number of training scenarios increases (e.g., $10$ and $15$), the generalization gap diminishes and becomes negligible. This suggests that exposure to a broader set of traffic realizations allows the learned policy to capture the underlying traffic interaction patterns rather than specializing to individual scenarios.

\subsection{Comparison against a Round-Robin Heuristic Baseline}

We additionally evaluate a round-robin heuristic alongside our learned policy. This heuristic assigns each aircraft one of the altitude levels as it approaches an intersection, granting access to one traffic stream via a round-robin scheduling policy across the incoming routes converging on that intersection. A route retains its assigned level until it empties or a maximum holding duration of $600$ seconds is reached, at which point in-progress aircraft are permitted to finish crossing before control rotates to the next waiting stream. Aircraft that have not yet been granted access have to hold outside the intersection boundary. Aircraft sharing a route additionally maintain a minimum separation. By construction, this heuristic guarantees that loss of separation events cannot occur.

\begin{table}[t]
\centering
\caption{Comparison of the round-robin heuristic and the trained RL policy. Values are mean $\pm$ std, pooled across
individual events, except LOS events, which is reported per scenario, and max
noise increase, which is a single maximum over scenarios.}
\label{tab:heuristic_vs_rl}
\begin{tabular}{lcc}
\toprule
Metric & Heuristic & RL \\
\midrule
LOS events         & $0.00 \pm 0.00$     & $4.30 \pm 2.85$     \\
Max noise increase (dB)         & $43.38$             & $38.85$             \\
Noise increase (dB)             & $24.45 \pm 6.28$    & $23.63 \pm 5.75$    \\
Travel time (s)                 & $635.39 \pm 197.83$ & $585.15 \pm 157.61$ \\
Midair halting time (s)         & $92.93 \pm 160.01$  & $0.00 \pm 0.00$     \\
Altitude Increases & $1.45 \pm 1.26$     & $10.63 \pm 3.34$    \\
\bottomrule
\end{tabular}
\end{table}

We evaluate both the round-robin heuristic and our noise and separation aware reinforcement learning policy ($\rho_{\text{sep}} = 0.95$, $\rho_{\text{noise}} = 0.05$) across the 20 scenarios used in our training and testing, and report the results in Table~\ref{tab:heuristic_vs_rl}. The heuristic reduces LOS events to zero, with a slight increase in noise relative to the RL policy. This safety guarantee comes at a cost of halting times. To guarantee separation, the heuristic must repeatedly cause aircraft to halt or hover in place. We therefore observe correspondingly higher travel times for the heuristic relative to the RL policy. We additionally observe that the number of altitude levels climbed per aircraft is substantially lower for the heuristic than for the RL policy. This is consistent with the heuristic's prescribed, event-driven altitude assignments, in contrast to the RL policy, which is free to adjust altitude at every timestep and consequently exhibits a higher frequency of climbing actions.

\section{Conclusion and Future Work}\label{sec:conclusion}

In this paper, we proposed an RL framework for managing noise impact, vertical separation, and energy consumption in UAM. By modeling decentralized aircraft behavior through a multi-agent RL formulation, our approach enables real-time altitude adjustments that balance competing operational objectives. Through extensive training and evaluation, we identified key tradeoffs among safety, noise impact, and energy efficiency. In particular, we found that policies can simultaneously achieve strong performance on separation and either noise or energy objectives. In contrast, noise and energy goals exhibit a stricter tradeoff, often requiring agents to prioritize one at the expense of the other. These results illustrate how multi-objective coordination strategies can be tuned to favor different operational goals, offering a pathway to adaptive, objective-driven management of future UAM network operations.

Looking ahead, several extensions may prove beneficial. First, while RL offers flexibility and adaptability, exploring its combination with optimization-based planning approaches could yield more holistic solutions with stronger interpretability guarantees. Second, our current model focuses on tactical decision-making at the vehicle level; integrating this with strategic separation assurance methods such as demand-capacity balancing may enable coordinated traffic flow management across larger urban networks. Together, these directions offer promising avenues for developing comprehensive and scalable solutions to the multifaceted challenges of UAM traffic management.

\section*{Appendix}

\subsection{Detailed Analysis on eVTOL Aircraft Energy Consumption}\label{sec:Appendix}

The analysis of eVTOL aircraft energy consumption during an UAM mission begins with a standard flight profile. Typically, a UAM flight profile consists of five segments: vertical takeoff, climb, cruise, descent, and vertical landing. In our study, there are five different cruising altitudes: 1,000 ft, 1,500 ft, 2,000 ft, 2,500 ft, and 3,000 ft. The total energy consumption for the mission is the sum of the energy used in all five segments. Below, we introduce an energy consumption model for each segment.

We begin by analyzing the two vertical segments. In the performance analysis of eVTOL aircraft, both vertical takeoff and landing can be approximated as hover in terms of energy consumption, which is the most energy-intensive part of the flight profile~\cite{kasliwal2019role}. Using classic momentum theory for helicopters~\cite{johnson2012helicopter}, we can calculate the hover power of an eVTOL aircraft as follows:
$P_{\text{hover}} = \frac{mg}{\eta_h} \sqrt{\frac{\delta}{2 \rho}}$
where $m$ (kg) is the aircraft's takeoff mass, $g = 9.81 ~\text{m}/\text{s}^2$ is the acceleration due to gravity, $\eta_h$ is the hover system efficiency, $\delta$ is the disk loading, and $\rho$ is the air density. Based on estimated data for a tilt-propeller configuration eVTOL aircraft~\cite{stoll2014conceptual,stoll2022development}, we use $\delta = 580~\text{N}/\text{m}^2$, $m = 1800~\text{kg}$, $\eta_h = 0.7$, resulting in:
\begin{equation}
    P_{\text{hover}} = \frac{1800~\text{kg} \cdot 9.81 ~\text{m}/\text{s}^2}{0.75} \sqrt{\frac{580~\text{N}/\text{m}^2}{2 \cdot 1.225~ \text{kg}/\text{m}^3}} = 362.3~\text{kW}
\end{equation}

The altitude AGL where vertical takeoff ends and vertical landing begins depends on the surrounding environment and topography of the vertiport. For this analysis, we assume that the aircraft takes 30 seconds to climb or descend 250 ft (76 m) during both hover segments. The energy consumption for vertical takeoff and landing is:
\begin{equation}
    E_{\text{hover}} = P_{\text{hover}} \cdot t_{\text{hover}} = 362.3~\text{kW} \cdot 60~\text{s} = 21.7~\text{MJ} 
\end{equation}
which remains constant for missions at all five altitudes.

Next, we examine the climb segment. We assume a small flight path angle of $\gamma = 10~\text{deg}$ and that, during this phase, the weight of the aircraft is roughly equal to the lift generated. The equation governing this segment is as follows:
\begin{equation}
    P_{\text{climb}} = \frac{TV_{\text{climb}}}{\eta_c} = \frac{V_{\text{climb}}}{\eta_c} (mg \sin{\gamma} + D) 
\end{equation}
where $\eta_c$ is the climb system efficiency. The drag $D$ is calculated using aerodynamic principles, expressed as follows:
\begin{equation}
\begin{aligned}
    D &= \frac{1}{2} \rho V_{\text{climb}}^2 S \left(C_{D_0} + K C_L^2\right)\\
    &= \frac{1}{2} \rho V_{\text{climb}}^2 S \left(C_{D_0} + \frac{1}{4 C_{D_0} (L/D)_{\text{max}}^2} \frac{(m g)^2}{(1/2 \rho V_{\text{climb}}^2 S)^2}\right)\\
    &= \frac{1}{2} \rho V_{\text{climb}}^2 S C_{D_0} + \frac{1}{4 C_{D_0} (L/D)_{\text{max}}^2} \frac{(m g)^2}{1/2 \rho V_{\text{climb}}^2 S}
\end{aligned}
\end{equation}
where $C_{D_0}$ is the zero-lift drag coefficient of the aircraft, $S$ is the reference area, and $(L/D)_{\text{max}}$ is the maximum lift-to-drag ratio. By combining the two equations above, we can express the climb power of an eVTOL aircraft as follows:
\begin{equation}
\begin{aligned}
    P_{\text{climb}} &= \frac{V_{\text{climb}}}{\eta_c} \large(mg \sin{\gamma}\\
    &+ \frac{1}{2} \rho V_{\text{climb}}^2 S C_{D_0} + \frac{1}{4 C_{D_0} (L/D)_{\text{max}}^2} \frac{(m g)^2}{1/2 \rho V_{\text{climb}}^2 S}\large)
\end{aligned}
\end{equation}

During the climb from a low altitude near the ground to cruising altitude, air density $\rho$ is not constant; it decreases with increasing altitude. Below 10,000 ft MSL, a linear model effectively represents this change in air density. Therefore, we use the mid-point air density $\rho_{\text{mid-h}}$ to approximate the air density during the climb. For instance, when the cruising altitude is $h = 2,000$ ft MSL, $\rho_{\text{mid-h}}$ corresponds to the air density at an altitude of $(2000 - 250)/2 + 250 = 1125$ ft. In this analysis, we assume a rate of climb (ROC) of 1,000 ft/min.

For the cruise segment, the power is modeled as follows:
\begin{equation}
\begin{aligned}
    &P_{\text{cruise}} = \frac{TV_{\text{cruise}}}{\eta_r} = \frac{V_{\text{cruise}}}{\eta_r} D\\ &= \frac{V_{\text{cruise}}}{\eta_r} \left(\frac{1}{2} \rho V_{\text{cruise}}^2 S C_{D_0} + \frac{1}{4 C_{D_0} (L/D)_{\text{max}}^2} \frac{(m g)^2}{1/2 \rho V_{\text{cruise}}^2 S}\right) 
    \end{aligned}
\end{equation}
where $\eta_r$ is the cruise system efficiency. Finally, for the descent segment, we assume that the engine power is reduced to 40\% of the power required for the cruise segment. In this study, we assume the following parameters: $\eta_c = 0.75$, $\eta_r = 0.8$, $(L/D)_{\text{max}} = 20$, $C_{D_0} = 0.03$, $S = 30~\text{m}^2$. With $V_{\text{climb}} = 1000/ (\sin(10~\text{deg}) \cdot 60) = 96~\text{ft/s}$, $V_{\text{cruise}} = 135~\text{ft/s}$, we can estimate the segment power consumptions.

With the power consumptions determined, we now calculate the energy consumption. We calculate the energy consumption for all 4 segments. First, the energy consumption for climb is
\begin{equation}
\begin{aligned}
    E_{\text{climb}} &= P_{\text{climb}} \cdot t_{\text{climb}}\\
    &= \frac{V_{\text{climb}}}{\eta_c} (mg \sin{\gamma} + \frac{1}{2} \rho_{\text{mid-h}} V_{\text{climb}}^2 S C_{D_0}\\ &+ \frac{1}{4 C_{D_0} (L/D)_{\text{max}}^2} \frac{(m g)^2}{1/2 \rho_{\text{mid-h}} V_{\text{climb}}^2 S}) \frac{(h-250)}{\text{ROC}}
\end{aligned}
\end{equation}

For the cruise segment, we first need to calculate the actual cruise distance. During the climb and descent segments, with a rate of 1,000 ft/min, the horizontal speed is given by $1000/\tan(10~\deg) = 5671$ ft/min = 94.5 ft/s. Thus, the combined distance traveled during the climb and descent is:
\begin{equation}
    d_{\text{climb+descent}} = 94.5~\text{ft/s} \cdot \frac{(h-250)\cdot 60}{1000} s \cdot 2 = 11.34(h-250)~\text{ft}
\end{equation}

The remaining distance for cruise is:
$d_{\text{cruise}} = d_{\text{total}} - d_{\text{climb+descent}}$. Therefore, the energy consumption during the cruise segment is:
\begin{equation}
\begin{aligned}
    E_{\text{cruise}} &= P_{\text{cruise}} \cdot t_{\text{cruise}}\\ &= \frac{V_{\text{cruise}}}{\eta_r} (\frac{1}{2} \rho V_{\text{cruise}}^2 S C_{D_0}\\ &+ \frac{1}{4 C_{D_0} (L/D)_{\text{max}}^2} \frac{(m g)^2}{1/2 \rho V_{\text{cruise}}^2 S}) \frac{d_{\text{cruise}}}{V_{\text{cruise}}}
\end{aligned}
\end{equation}

The energy consumption during the descent segment is:
\begin{equation}
\begin{aligned}
    E_{\text{descent}} &= 0.4 P_{\text{cruise}} \cdot t_{\text{descent}}\\ &= \frac{0.4 V_{\text{cruise}}}{\eta_r} (\frac{1}{2} \rho V_{\text{cruise}}^2 S C_{D_0}\\ &+ \frac{1}{4 C_{D_0} (L/D)_{\text{max}}^2} \frac{(m g)^2}{1/2 \rho V_{\text{cruise}}^2 S}) \frac{(h-250)}{\text{ROC}}
\end{aligned}
\end{equation}

With all segment energy consumption models considered, the total energy consumption for the entire mission is:
\begin{equation}
    E_{\text{total}} = E_{\text{hover}} + E_{\text{climb}} + E_{\text{cruise}} + E_{\text{descent}}
\end{equation}

\begin{figure}[hbt!]
     \centering
\includegraphics[width=0.37\textwidth]{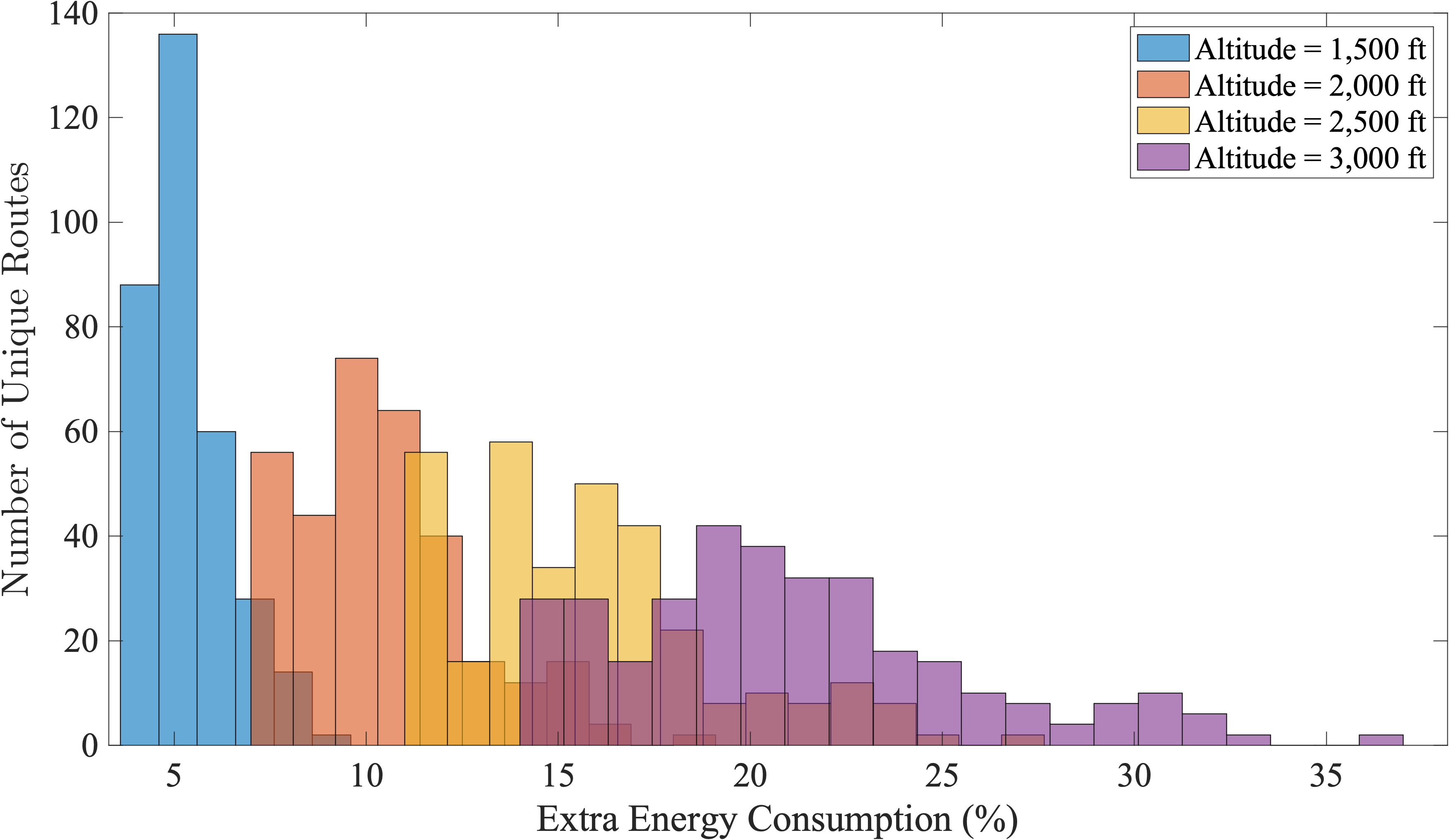}
     \caption{Histogram on extra energy consumption distributions for the Austin UAM network}
     \label{fig:EnergyAltitude}
\end{figure}

Within the operational range of UAM,  cruising at a higher altitude leads to increased total energy consumption $E_{\text{total}}$. This is due to the longer duration spent in the climb segment. Although cruising at a higher altitude reduces drag and consequently decreases energy consumption during cruise $E_{\text{cruise}}$, the benefits are outweighed by the extended climb.

However, the actual percentage increase in energy consumption from cruising at a higher altitude depends on the mission distance. This additional energy consumption increases as the distance decreases. We examined the extent of this additional energy consumption across the Austin UAM network. Using the energy consumption at a cruising altitude of 1,000 ft as a baseline, Figure~\ref{fig:EnergyAltitude} shows the alternative energy consumptions at four higher altitudes for all possible routes within the Austin UAM network. Analyzing the mean results for all routes, cruising at altitudes of 1,500 ft, 2,000 ft, 2,500 ft, and 3,000 ft lead to 5\%, 10\%, 15\%, and 20\% extra energy consumption, respectively.

\bibliographystyle{IEEEtran}
\bibliography{sample}
\begin{IEEEbiography}[{\includegraphics[width=1in,height=1.25in,clip,keepaspectratio]{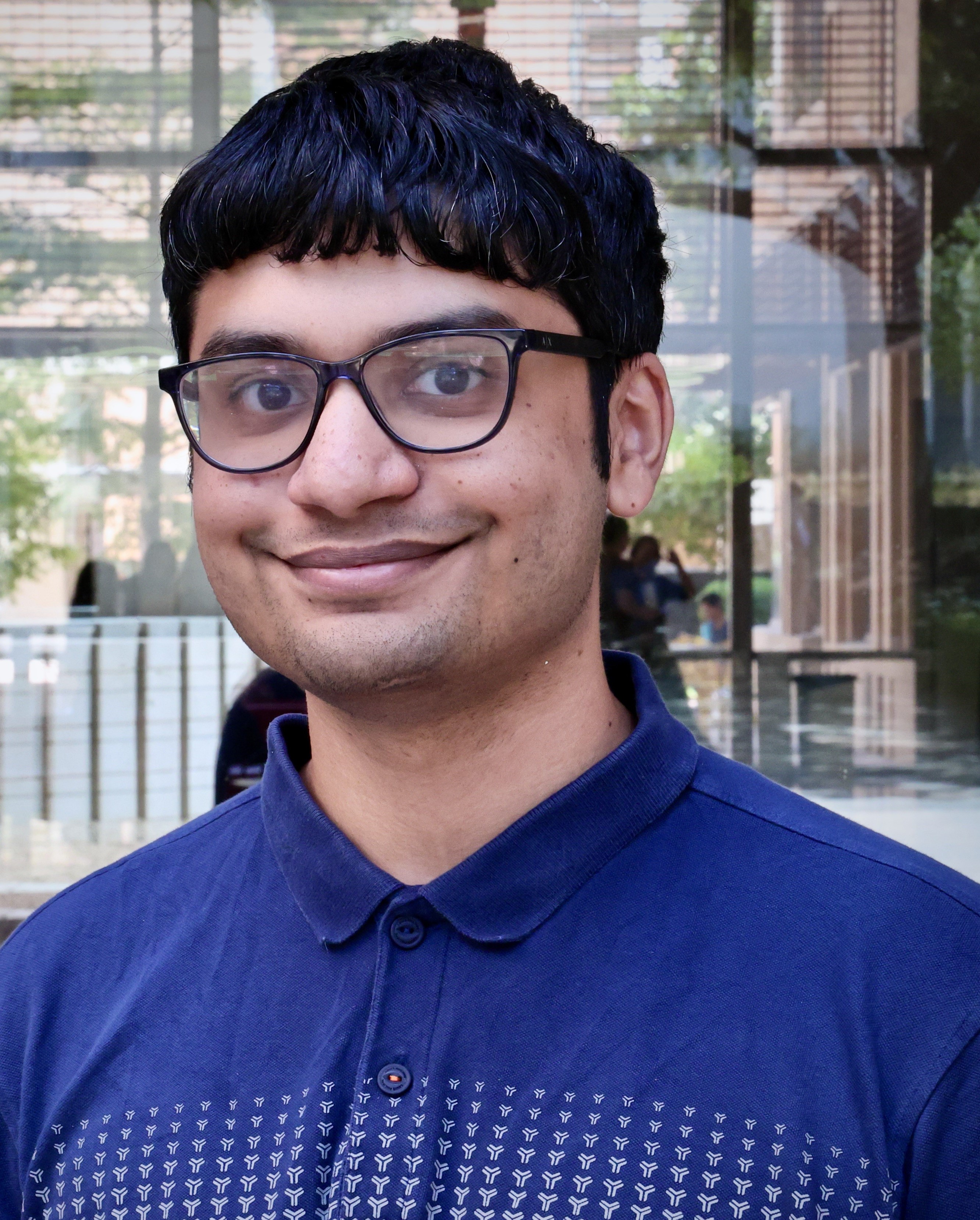}}]{Surya Murthy} is a PhD student in the Electrical and Computer Engineering department at the University of Texas at Austin. He received his B.S. degree in Computer Engineering from the University of Illinois at Urbana-Champaign. His research interests are multi-agent systems, multi-objective decision-making, and human-agent collaboration.
\end{IEEEbiography}
\begin{IEEEbiography}[{\includegraphics[width=1in,height=1.25in,clip,keepaspectratio]{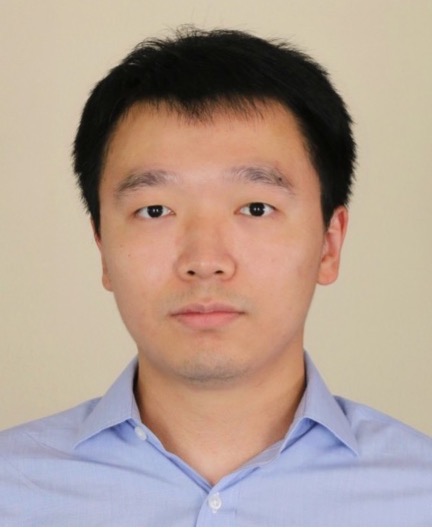}}]{Zhenyu Gao} a postdoctoral fellow at The University of Texas at Austin. He obtained his Ph.D. in Aerospace Engineering and an M.S. in Operations Research from the Georgia Institute of Technology. His research interests encompass systems design and optimization, data-driven methods, and AI for aerospace applications. 
\end{IEEEbiography}
\begin{IEEEbiography}[{\includegraphics[width=1in,height=1.25in,clip,keepaspectratio]{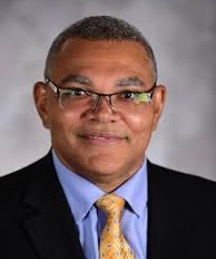}}]{John-Paul Clarke} is a Professor at The University of Texas at Austin, where he holds the Ernest Cockrell Jr. Memorial Chair in Engineering. He received the S.B., S.M., and Sc.D. degrees from MIT. He is a Fellow of the AIAA and the Royal Aeronautical Society, and a Member of the National Academy of Engineering. His research focuses on the development and use of stochastic models and optimization algorithms to improve the efficiency and robustness of complex systems, with a particular focus on aviation.
\end{IEEEbiography}
\begin{IEEEbiography}[{\includegraphics[width=1in,height=1.25in,clip,keepaspectratio]{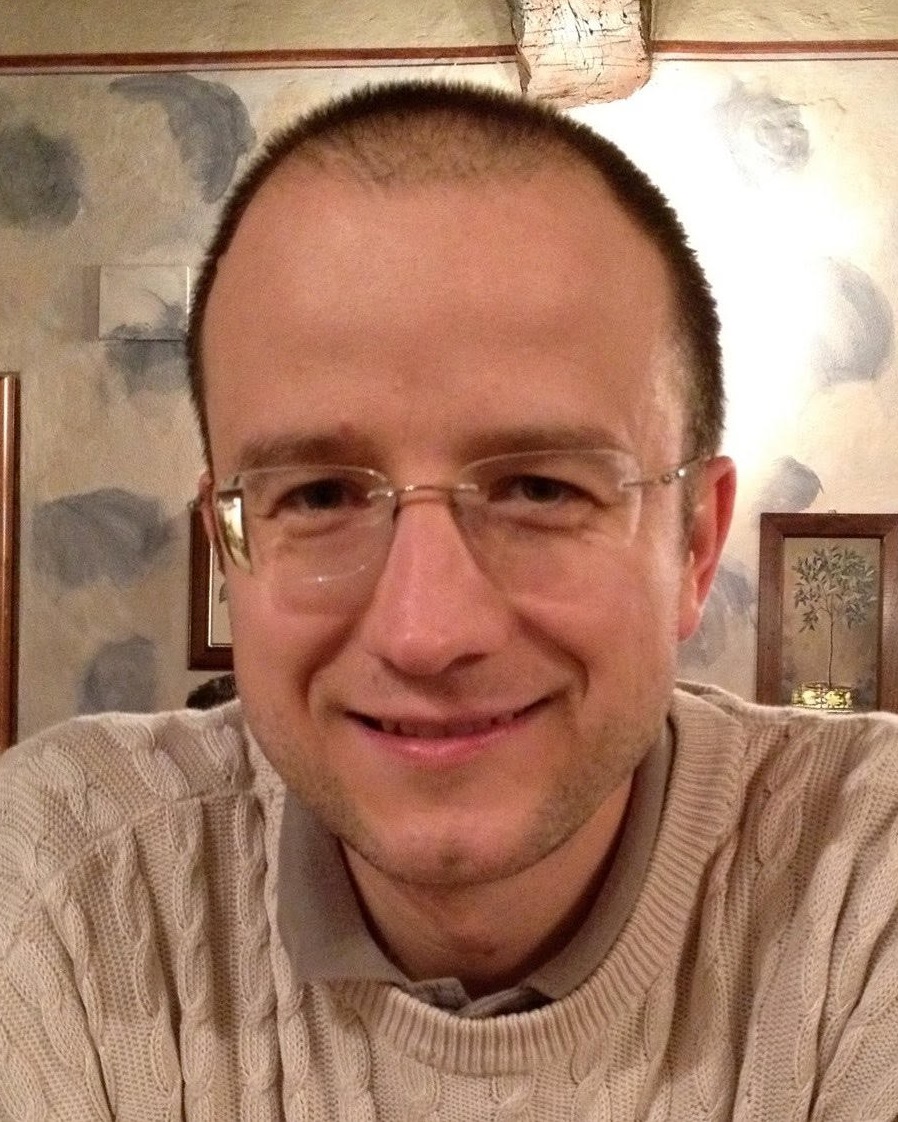}}]{Ufuk Topcu} received the Ph.D. degree from the University of California at Berkeley, Berkeley, CA, USA, in 2008. He joined the Department of Aerospace Engineering, University of Texas at Austin, Austin, TX, USA, in Fall 2015. His research focuses on the theoretical, algorithmic and computational aspects of design and verification of autonomous systems through novel connections between formal methods, learning theory, and controls.
\end{IEEEbiography}
\end{document}